# Spatial-Knowledge-Graph-Grounded LLM Agents for Neighborhood Livability Evaluation

Haiyan Hao
School of Humanities and Social Science
The Chinese University of Hong Kong, Shenzhen
haohaiyan@cuhk.edu.cn
August 27, 2026
(Web App demonstration: https://www.urbanresilienceai.com/activity_simulator/)

## Abstract

Neighborhood livability is commonly assessed with static built-environment indicators, such as facility proximity, street connectivity, and access to public space. These measures describe available opportunities but do not directly represent how residents with different mobility capacities, household roles, schedules, and care responsibilities experience the neighborhood. This paper presents a prototype framework that uses a spatial knowledge graph (KG) and large language models (LLMs) to generate and revise household schedules, followed by rule-based feasibility checking and GIS-based network materialization. The spatial KG integrates residents, residences, facilities, neighborhood context, and sampled road hubs; Graph-RAG retrieves each household's nearby spatial context, including candidate POIs and approximate walking times, for the scheduling LLM. The LLM produces structured household schedules, while rules are used for lightweight repairs and auditable feasibility checks. The LLM then revises schedules in response to identified feasibility issues. A routing module derives the actual travel paths, travel times, modes, and event histories from the road network. The resulting events support synthetic resident-agent interviews about daily convenience, travel burden, activity feasibility, and household coordination. A prototype demonstration in a Shenzhen neighborhood shows that nominal facility availability does not necessarily imply convenient access: residents with limited mobility and households with care responsibilities experience greater travel and coordination burdens. The framework offers an auditable way to connect spatial opportunity, household activity constraints, and resident-specific livability interpretation, while keeping simulated experience distinct from observed perception.

## 1. Introduction

Neighborhood livability concerns the extent to which a residential environment supports everyday life, wellbeing, and participation. It is commonly evaluated through built-environment indicators such as access to services, public open space, transportation hubs, walkability, and safety. These indicators provide an important basis for planning comparison. For example, the 15-minute-city literature emphasizes proximity to daily service destinations as a pathway toward more sustainable and equitable urban life (Moreno et al., 2021), while broader livability frameworks identify access to health and social services, food, transport, open space, safety, and social cohesion as key domains (Badland et al., 2014). However, such measures usually describe what a neighborhood provides, rather than how different residents experience those provisions in daily life.

This distinction is important because the same spatial environment may impose very different burdens on different residents. A facility that is nearby for a healthy adult may be difficult to reach for an older resident with slow walking speed. A neighborhood with schools, parks, grocery stores, and clinics may still be inconvenient for a household coordinating work schedules, school pick-up, childcare, and shopping when the required destinations lie in different directions and cannot be combined into a coherent activity chain. Conventional accessibility measures often evaluate aggregate spatial opportunities across residents using isolated origin-destination relationships, with origins often set to residential addresses. They are less able to capture realistic travel behaviors, such as activity chains linking school escort, work, and shopping; temporal constraints imposed by fixed commitments; the coordination of shared tasks among household members; and age- or disability-related mobility limitations. These concerns are central to activity-based and person-based accessibility research, which emphasizes that access depends not only on the location of opportunities but also on individual capabilities, schedules, and feasible space-time paths (Dong et al., 2006; Kwan, 1998; Neutens et al., 2008).

Recent advances in LLMs provide new possibilities for representing heterogeneous resident agents. LLM-based agents can interpret rich persona descriptions, reason about household-level trade-offs, and generate plausible schedules. However, when such processes are not grounded in explicit spatial evidence, plausible outputs may be inconsistent with actual destinations, routes, and travel conditions, limiting their credibility for spatial simulation and evaluation.

An important gap therefore remains between spatial accessibility measurement and resident-specific livability evaluation. Conventional spatial indicators provide limited insight into how individuals and households with different schedules, mobility capacities, and care responsibilities organize everyday activities within the same neighborhood. Meanwhile, LLM agents can represent such heterogeneity but often lack structured spatial context and an auditable connection between generated behavior and subsequent evaluation. To address this gap, we develop a spatial-KG-grounded LLM-agent framework that simulates household activity-mobility schedules and uses the resulting event histories to support evidence-grounded, post-simulation resident interviews.

This study makes three contributions. First, it advances neighborhood evaluation from place-based static accessibility measurement toward personalized assessment by representing how personal characteristics and household circumstances shape the feasibility and burden of everyday activities. Second, it introduces evidence-grounded post-simulation interviews that translate activity-mobility events into interpretable resident-specific evaluations while preserving links to the underlying evidence. Third, it structures residents, facilities, and spatial relationships through a spatial KG, providing LLM agents with a geographically grounded and traceable context for household schedule generation. Together, these contributions connect spatial opportunity, heterogeneous daily behavior, and subjective interpretation within an auditable evaluation framework.

## 2. Related Work

### 2.1 Neighborhood Livability and Accessibility

Neighborhood livability is commonly understood as the capacity of a place to support residents' everyday needs, wellbeing, and participation in community life (Badland et al., 2014; Badland & Pearce, 2019). It is a broad and contested concept rather than a single measurable attribute. Commonly recognized domains include transportation, access to local food and other goods, health and social services, public open space, housing, safety, natural environment, leisure and culture, and social cohesion (Badland et al., 2014). In empirical research, these domains are frequently operationalized through objective spatial indicators, such as amenity density and diversity, distance to services, public-transport provision, walkability, green-space access, and environmental exposure (Zhang & Kwan, 2025). However, these indicators generally assume that access can be inferred from spatial opportunity alone. They rarely represent whether residents can combine multiple activities within daily schedules or whether household members can coordinate mobility under care, work, and school constraints.

Accessibility to facilities that support everyday needs is a core component of neighborhood livability (Hao & Wang, 2022; Wu et al., 2025). Conventional measures include cumulative-opportunity indicators, which count destinations reachable within a specified travel-time threshold; gravity-based indicators, which weight opportunities by travel impedance; and proximity measures, which report the distance or time to the nearest relevant destination (Geurs & van Wee, 2004). These measures are widely used to diagnose inequalities in service provision and to evaluate land-use and transport strategies.

Accessibility indicators also underpin the X-minute-city agenda, which seeks to ensure that essential daily needs can be reached through short active-travel trips from home (Moreno et al., 2021; Logan et al., 2022; Khavarian-Garmsir et al., 2023). This agenda has stimulated extensive GIS-based evaluation. Typical studies identify essential facilities, estimate walking, cycling, or public-transport travel times, and summarize the share of residents or locations that meet a predefined threshold. Recent work has expanded these evaluations by incorporating multiple modes, population coverage, and cross-city comparison (e.g., Logan et al., 2022; Wang et al., 2025). Human-mobility research further suggests that local amenity access is associated with the extent to which people conduct activities near home (Abbiasov et al., 2024). Together, these studies establish that built-environment conditions shape the spatial opportunities available to residents.

### 2.2 Activity-Based and Household-Aware Accessibility

The opportunities available to residents are often not used through isolated trips. Daily mobility commonly involves activity chains, e.g., a parent may travel from work to collect a child from school, stop at a grocery store, and then return home. Whether such a chain is feasible depends not only on the proximity of each destination to home, but also on the spatial relationship between destinations, their opening times, the sequence of activities, and the time available between commitments. Two facilities may each be close to home but located in opposite directions, making them difficult to combine efficiently within the same trip. Activity-based accessibility therefore shifts attention from access to a single destination toward the set, sequence, and timing of activities that a person can undertake (Dong et al., 2006).

Time-geographic approaches similarly emphasize that accessibility is constrained by the space-time paths individuals can feasibly traverse. Kwan (1998) showed that person-based space-time measures capture interpersonal differences that are obscured by place-based or aggregate accessibility measures. A resident's feasible activity space depends on mobility resources, fixed anchors such as work or school, travel time, available time, and the temporal availability of destinations. These constraints are not merely technical details: they are mechanisms through which gender, age, employment status, disability, and care responsibilities can produce unequal access to urban opportunities.

Household interactions further complicate accessibility. Many everyday activities are coordinated across household members: adults may escort children, accompany older relatives, divide shopping and care tasks, or modify their own schedules to accommodate another member's obligations. Recker et al. (2001) demonstrated that household interactions and time-space constraints affect the accessibility benefits generated by travel decisions. Neutens et al. (2008) similarly showed that joint accessibility depends on the schedules, travel times, and availability constraints of all participants in a shared activity. A neighborhood may therefore be convenient for an independent adult while creating substantial coordination burdens for a household with children, older adults, or members with limited mobility.

These insights expose a limitation in dominant GIS-based livability and accessibility assessments. Location-based indicators can reveal the distribution of opportunities, but they generally do not indicate whether those opportunities can be combined into feasible activity chains, coordinated across household members, or completed within realistic time windows. Conversely, activity-based and joint-accessibility models offer more behaviorally grounded accounts of constraint, but are often designed to estimate accessibility or travel demand rather than to represent how accumulated daily experiences shape residents' evaluations of neighborhood livability. There remains a need for approaches that connect spatial opportunities, daily activity constraints, and household coordination to interpretable, resident-specific assessments of neighborhood life.

### 2.3 LLM Agents in Urban Simulation

LLMs have created new possibilities for representing agents that can generate plans, use contextual information, and communicate explanations in natural language (Hao et al., 2024). Park et al. (2023) introduced the generative-agent paradigm, in which agents use stored observations, reflection, and planning to generate believable social behavior in a simulated environment. Since then, LLM-based agents have been applied to urban planning, mobility simulation, and urban digital-twin research. For example, Zhou et al. (2024) used LLM agents to simulate resident participation in land-use planning, while Song et al. (2025) used LLMs to generate heterogeneous population profiles, daily locations, and mobility patterns at urban scale. Recent mobility-focused work has sought to equip agents with more elaborate cognitive structures. Liu et al. (2026) proposed GATSim, which integrates LLM-based planning, memory, reaction, and reflection into an urban mobility simulation framework. Other work has integrated LLM-based replanning agents into conventional transport simulation environments, using verification mechanisms and domain tools to constrain agent decisions (Patwary et al., 2026).

These studies suggest that LLMs can enrich conventional rule-based systems by representing preferences, adaptive decisions, and natural-language interaction that are difficult to encode exhaustively by hand. Nevertheless, the relationship between activity generation and spatial context remains insufficiently resolved. In many modular simulation workflows, LLMs first generate semantic activities or approximate schedules from persona and temporal information, after which specific destinations, routes, and travel times are assigned by GIS or transport-simulation modules (Song et al., 2025; Lämmer et al., 2026). Although this separation simplifies model implementation, activity scheduling is inherently spatial. Whether an activity can be fitted into a daily schedule depends on where suitable destinations are located, how they relate to preceding and subsequent destinations, and whether the resulting travel can be completed within available time windows. A schedule may therefore appear temporally and behaviorally plausible but become inefficient or infeasible after spatial realization (Santos et al., 2026).

Graph-based retrieval offers a potential means of bringing spatial context into activity generation. GraphRAG retrieves relevant entities together with their relationships, enabling LLMs to use structured external knowledge during generation (Edge et al., 2024). In mobility applications, a spatial KG can encode spatial relationships among residents, residences, and other spatial features (e.g., roadways and facilities). A household-centered ego-subgraph can then be retrieved from the broader spatial KG, providing the LLM with local context, such as destination availability and approximate travel burdens, when generating activity schedules.

Despite these developments, limited attention has been given to how spatially grounded LLM agents can support neighborhood livability evaluation. Existing studies have primarily focused on generating mobility behavior, reproducing travel patterns, or enabling adaptive replanning. Less is known about how household-level schedules and their network-derived consequences can be translated into resident-specific assessments of neighborhood conditions. Addressing this gap requires connecting spatially informed activity generation with traceable mobility outcomes and evidence-grounded interpretation of residents' simulated daily experiences.

## 3. Framework and Prototype System

### 3.1 System Workflow and Spatial Knowledge Graph

The prototype comprises four connected modules: (1) Spatial Setup and KG Construction; (2) Household and Persona Configuration; (3) Iterative Scheduling & Network Materialization; and (4) Inspection and Interviews (**Figure 1**). Together, these modules connect neighborhood spatial opportunities with simulated household activities, mobility outcomes, and resident-specific evaluations.

First, the Spatial Setup and KG Construction module defines the local environment as $\mathcal{E} = (G, F, B)$, where $B$ is the user-defined residential boundary, $G = (V, L)$ is the routable road graph, and $F$ is the facility set. Roads and facilities are retrieved from OpenStreetMap within a 3-km buffer around $B$. Local GIS layers can be substituted or added where more detailed local data are available. These spatial data are organized into a spatial KG, denoted by $\mathcal{K}$. Second, the Household and Persona Configuration module represents the resident population as $\mathcal{A} = \{a_1, a_2, \ldots, a_n\}$ and organizes agents into households $\mathcal{H} = H_1, \ldots H_J$ with shared activities, constraints, and care responsibilities. Third, for each household $H_j$, the scheduling module retrieves a household-centered ego-subgraph $\mathcal{K}_j \subseteq \mathcal{K}$, generates and reviews a household schedule through an iterative LLM-rule process, and materializes the accepted schedule on $G$, which returns an event set $\mathcal{Z}$. Finally, the inspection and interview module presents the resulting events through maps, timelines, and aggregate indicators and uses them as evidence for resident-agent interviews.

The prototype is implemented as a web-based system with a React/TypeScript front end and a Python FastAPI back end. The back end uses Pydantic data schemas, Shapely for spatial operations, NetworkX for routing, and an LLM API for schedule generation and revision. REST APIs transmit user configurations, trigger spatial retrieval and simulation, and return KG summaries, schedule and event records, aggregate metrics, and interview responses. The web application is publicly available at: https://www.urbanresilienceai.com/activity_simulator/.

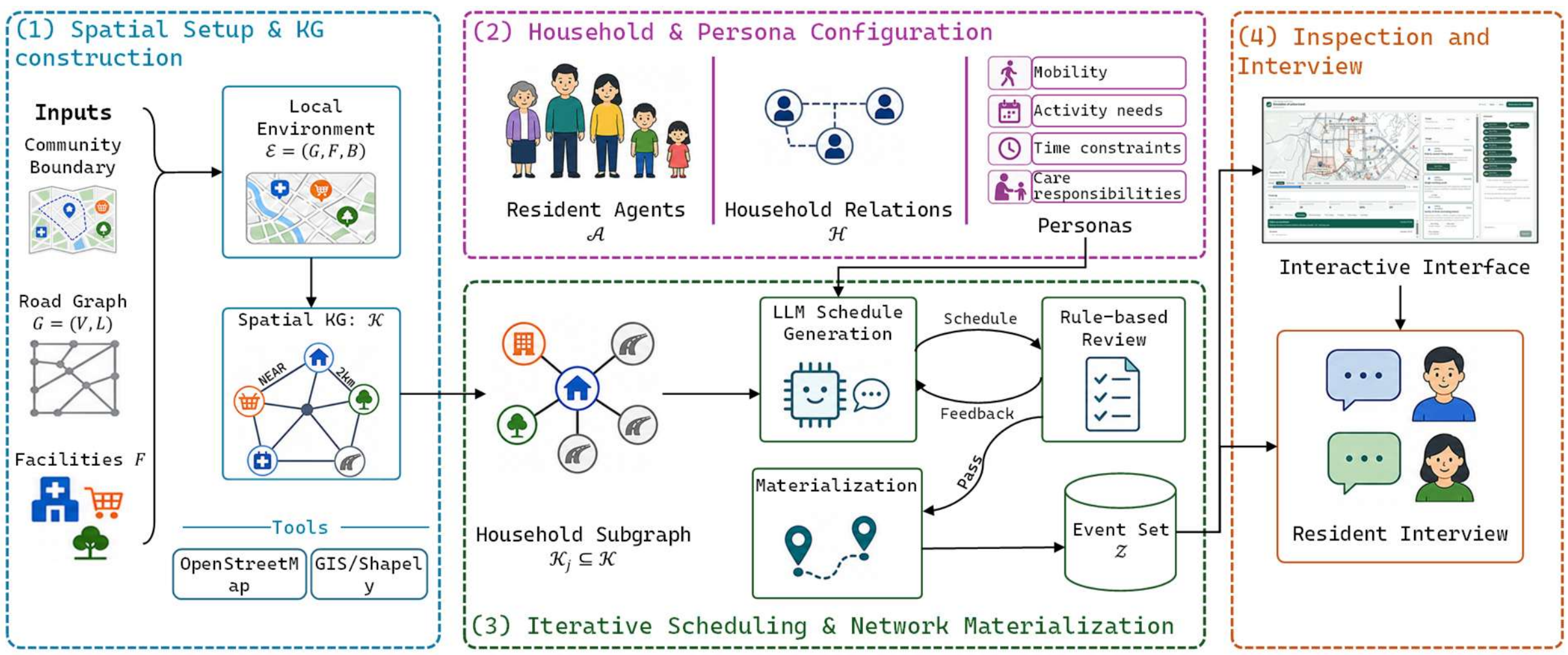

**Figure 1.** System architecture and information flow.

### 3.2 Household and Persona Configuration

Each resident agent is initialized from a structured persona specifying demographic characteristics, household role, mobility capacity, preferred travel modes, activity needs, behavioral constraints, fixed spatial anchors, and temporal commitments. Household and care attributes further describe relationships among agents, including caregiving, escort, and accompaniment responsibilities.

These attributes constrain household schedule generation. Mobility capacity affects travel speed, while activity needs determine the facility categories relevant to each resident. Behavioral and temporal constraints regulate when and how activities may occur, and fixed anchors specify externally determined commitments such as work and school. Household relationships determine whether an activity can be undertaken independently or requires the participation of another member. Household membership does not imply that all members travel together. For each activity, the traveling party is determined by the activity type, care requirements, and the availability of eligible household members. For example, a child cannot initiate an independent trip and must be accompanied by an available adult when required.

The prototype includes five predefined individual and household profiles representing population groups relevant to accessibility and livability research (**Table 1**). Users may select these profiles for standardized comparisons or create custom personas for scenario-specific analyses.

**Table 1.** Built-in persona library

| Profile | Type | Key configuration | Analytical rationale | Citation |
|---|---|---|---|---|
| Independently living older adult | Individual | Slow walking; grocery, clinic, pharmacy, park, and social needs; avoids | Examines the burden of reduced walking speed and dependence on nearby daily | WHO (2007); Moreno et al. (2021) |

| | | late-night trips | and health services | |
|---|---|---|---|---|
| Young professional | Individual | Work anchor; walking and bus preferences; fragmented weekday time | Represents a commuter whose local opportunities must fit around a fixed work schedule | Pozoukidou and Chatziyiannaki (2021); Neutens (2015) |
| Dual-earner household with infant | Household | Two adults and an infant; childcare, stroller-related constraints, clinic and grocery needs | Represents care-related time-space constraints and household coordination | Kwan (1999); Moreno et al. (2021) |
| Household with school-age child | Household | Parents and child; school anchor, escorting, after-school and grocery activities | Examines school travel, pick-up responsibilities, and multipurpose activity chains | McMillan (2005); Recker et al. (2001) |
| Mobility-constrained adult | Individual | Slow walking; short and simple trips; clinic, pharmacy, grocery, and park needs | Examines unequal access resulting from limited mobility capacity | Neutens (2015); WHO (2007) |

### 3.3 Spatial-KG-Grounded LLM Scheduling and Network Materialization

The prototype simulates a seven-day period with time represented in minutes. Scheduling is performed at the household level through an LLM-first, rule-supported process: the LLM generates and revises household schedules, deterministic rules perform limited repairs and feasibility checks, and network routing materializes accepted schedules as activity–mobility events.

#### *Spatial KG Construction and Retrieval*

The spatial KG is denoted by $\mathcal{K} = (N, R)$, where $N$ and $R$ contain the node and relation sets summarized in **Table 2** and **Table 3**. It is constructed from the local environment $\mathcal{E}$, resident-agent set $\mathcal{A}$, and household set $\mathcal{H}$. The KG represents neighborhoods, residents, households, residences, facilities, activity categories, and sampled road hubs, together with their spatial and household relationships.

For each residence, facilities are grouped by category and ranked by haversine distance. The KG retains up to ten facilities per category within 3 km through `NEAR` relations. Each relation stores the straight-line distance and an approximate walking time calculated using a reference speed of 70 m/min. These values support subgraph retrieval and preliminary schedule planning, whereas final routes and travel times are calculated on the complete road network $G$.

**Table 2.** Node schema of the spatial knowledge graph

| Type | Key attributes | Description |
|---|---|---|
| Neighborhood | id, name, lat, lon, radiusMeters | Represents the user-defined neighborhood and |

| | | its spatial extent. |
|---|---|---|
| Person | id, name, role, age, summary, mobility, preferredModes, profileId | Represents an individual resident agent and the attributes relevant to activity and mobility decisions. |
| Household | id, name, profileId | Represents a household containing one or more resident agents. |
| Residence | id, name, lat, lon, profileId | Represents the residential location shared by members of a household. |
| Facility | id, name, category, lat, lon, osmId | Represents a potential activity destination obtained from the spatial data. |
| Category | id, name | Represents an activity-relevant facility category, such as grocery, clinic, pharmacy, or park. |
| RoadHub | id, name, lat, lon, degree | Represents a sampled high-connectivity node from the local road graph. |

**Table 3.** Relation schema of the spatial knowledge graph

| Type | Source node | Target node | Key attributes | Description |
|---|---|---|---|---|
| LIVES_AT | Person | Residence | - | Links a resident agent to their residential location. |
| MEMBER_OF | Person | Household | - | Links a resident agent to the household to which they belong. |
| IN_AREA | Residence, Facility, or RoadHub | Neighborhood | - | Indicates that a spatial entity is located within the modeled neighborhood context. |
| OF_CATEGORY | Facility | Category | - | Assigns a facility to an activity-relevant category. |
| NEAR | Residence | Facility | meters, walkMin | Links a residence to a nearby facility and stores straight-line distance and approximate walking time. |
| ROAD_LINK | RoadHub | RoadHub | meters, walkMin | Represents simplified connectivity between sampled road hubs and stores approximate separation and walking time. |

For each household $H_j$, Graph-RAG retrieves a household-centered subgraph $\mathcal{K}_j \subseteq \mathcal{K}$. Retrieval begins from the household members, follows their LIVES_AT relations to the shared residence $b_j$, and then follows NEAR relations to candidate facilities. For each activity category $c$, up to $q = 10$ facilities are selected:

$$F_{j,c}^{(q)} = \mathrm{Top}_q\left(\{f \in F_c : (b_j, f) \in R_{\mathrm{NEAR}}\}; \hat{\tau}_{b_j f}\right).$$

where candiates are ranked by their approximate walking time from $b_j$, $F_c$ is the set of facilities of category $c$, and $F_{j,c}^{(q)}$ is the corresponding candidate set retrieved for household $H_j$.

The resulting subgraph used in LLM is:

$$\mathcal{K}_j = \mathcal{K}\left[H_j \cup \{b_j\} \cup \bigcup_c F_{j,c}^{(q)}\right].$$

The retrieved subgraph provides the scheduling LLM with household attributes, residence information, valid POI identifiers, facility categories, and approximate walking times. This bounded context allows destination availability and travel burden to inform schedule generation while preventing the LLM from inventing unsupported facilities.

***Household Schedule Generation and Review***

The scheduling LLM receives the household personas, ego-subgraph $\mathcal{K}_j$, simulation horizon $D$, scheduling requirements, and output schema. It returns a household schedule in tab-separated value (TSV) format, which is more compact than JSON and requires fewer output tokens.

Each schedule row records the day, start and end times, activity type, lead member, companions, destination POI, activity label, chaining status, and optional travel mode. The generated schedule then enters an iterative rule-based review. In each review round $r$, the rule layer first applies conservative repairs, including removing exact duplicates, consolidating joint activities, normalizing school escort records, joining adjacent work segments, and inserting unambiguous return-home movements. It then expands the schedule into a complete 24-hour timeline for each resident agent. Travel, activities, and unscheduled intervals are resolved into continuous person-level trajectories, making overlaps, inconsistent origins, and sudden location changes easier to detect. The remaining unresolved issues forms a diagnostic set $\Delta_j^{(r)}$.

If $\Delta_j^{(r)} \neq \emptyset$, the LLM receives the current TSV schedule, $\Delta_j^{(r)}$, household context, and ego-subgraph and returns a targeted revision. The revised schedule re-enters the same repair and diagnostic process. This iterative process continues until $\Delta_j^{(r)} = \emptyset$ or the number of round $r$ reaches a predefined threshold $r_{\max}$, at which the simulation reports a scheduling failure rather than returning an incomplete household schedule.

The principal feasibility checks are:

a) **Valid time and duration**. Each activity must start and end within the same simulated day, with $0 \leq t_s^s < t_s^e \leq 1440$. Its duration must accommodate the approximate travel time and the duration associated with its activity type.

b) **Participant exclusivity**. The lead member and all companions are treated as participants in an activity. Their scheduled intervals are compared across the household to ensure that no person is assigned to two overlapping activities.
c) **Spatiotemporal continuity**. Consecutive activities must allow sufficient time for movement between their locations. The schedule cannot place an agent or companion at a new destination while that person remains occupied elsewhere. This prevents temporal overlap and “teleportation”.
d) **Mandatory anchors and weekly patterns**. Work and school commitments are enforced on applicable weekdays rather than automatically repeated across all seven days. Weekend work requires an explicit persona constraint, such as shift work.
e) **Household care constraints**. Infants cannot travel independently, and children under 12 undertaking non-school activities must be accompanied by an eligible adult. A household member cannot escort or collect a child while simultaneously participating in another activity.
f) **Joint-activity consistency.** A joint activity is represented by one lead row and its companion set, rather than duplicated as separate simultaneous rows for multiple members.
g) **Daily continuity and return home.** Unexplained temporal gaps between successive out-of-home activities are flagged, and each participating resident must end the simulated day at home.
h) **Destination validity.** Non-anchor POIs must belong to the candidate set retrieved in $\mathcal{K}_j$, and activity types must belong to the supported activity taxonomy.

*Network Materialization and Event Construction*

After acceptance, activities are ordered by day and departure time. For each activity, the materialization module identifies the traveling party, current origin, and destination. Work and school activities use fixed anchors, whereas other activities use POIs selected from $\mathcal{K}_j$.

For each scheduled trip, the origin and destination are snapped to the nearest road-network node, and NetworkX is used to calculate a distance-weighted shortest path. Travel mode is assigned through a rule-based procedure that considers the resident’s available modes and the estimated walking time. Walking is used for shorter trips, whereas longer trips may be assigned to transit when a feasible connection through nearby transit stops can be constructed; residents with access to a car may be assigned the driving mode. Network distance and travel time are then derived from the resulting path and the mode-specific reference speed.

The origin of each movement is inherited from the preceding realized event, allowing activity chains to be represented as destination-to-destination movements. A member-level clock delays activities when necessary, maintains the latest location of each participant, inserts missing return-home movements, and performs a final participant-overlap check.

Each materialized event is represented as

$$z = (a_i, C, d, t^s, t^e, \alpha, o, f, p, m, \delta, \tau, y, \varepsilon),$$

where $a_i \in \mathcal{A}$ is the lead resident and $C \subseteq H_j \setminus \{a_i\}$ is the set of accompanying household members; $d$, $t^s$, $t^e$, and $\alpha$ denote the simulated day, start and end times, and activity type; $o$, $f$, $p$, $m$ denote the origin, destination, network path, and realized travel mode; $\delta$ and $\tau$ denote the network distance and travel time. The binary variable $y$ indicates whether the event was successfully materialized; and $\varepsilon$ records the failure reason when $y = 0$. The complete output is the event set for all households:

$$\mathcal{Z} = \{z_1, z_2, \dots, z_{|Z|}\}.$$

***Provenance, Failure Handling, and Reproducibility***

Each run retains the identifier and metadata of $\mathcal{K}$, the retrieved household contexts $\mathcal{K}_j$, initial and revised schedules, rule-patch notes, diagnostic sets $\Delta_j^{(r)}$, prompt and model metadata, and the materialized event set $\mathcal{Z}$. This provenance distinguishes retrieved spatial facts, LLM-generated decisions, deterministic repairs, diagnostic feedback, and network-derived outcomes. **Algorithm 1** summarizes the prototype workflow.

The API exposes $\mathcal{K}$, household schedules, diagnostic provenance, $\mathcal{Z}$, and the resulting metrics $\mathcal{M}$ for KG inspection, timeline and map visualization, and resident interviews. For larger-scale applications, the event records can be stored in a spatially enabled relational database.

**Algorithm 1.** Spatial-KG-grounded household schedule generation and network materialization

Input:
local environment $\mathcal{E} = (G, F, B)$; resident-agent set $\mathcal{A}$; household sets $\mathcal{H}$
simulation horizon $D = \{0, \ldots, 6\}$; maximum revision rounds $Rmax$

Output:
spatial KG $\mathcal{K}$; accepted household schedules $S$; event set $\mathcal{Z}$; weekly metrics $\mathcal{M}$

1: $\mathcal{K} \leftarrow$ BuildSpatialKG($\mathcal{E}$, A, $\mathcal{H}$)
2: $\mathcal{Z} \leftarrow \emptyset$; $S \leftarrow \emptyset$
3: for each household $H_j \in \mathcal{H}$ do
4: $\quad \mathcal{K}_j \leftarrow$ RetrieveEgoSubgraph($\mathcal{K}, H_j$)
5: $\quad S_j^{(0)} \leftarrow$ LLMGenerate($H_j, \mathcal{K}_j, D$)
6: $\quad$ accepted ← False
7: $\quad$ for $r \leftarrow 0$ to $Rmax$ do
8: $\quad\quad \tilde{S}_j^{(r)} \leftarrow$ LightPatch($S_j^{(r)}$)
9: $\quad\quad \Delta_j^{(r)} \leftarrow$ RuleCheck($\tilde{S}_j^{(r)}, H_j, \mathcal{K}_j, D$)
10: $\quad\quad$ if $\Delta_j^{(r)} = \emptyset$ then
11: $\quad\quad\quad S_j \leftarrow \tilde{S}_j^{(r)}$
12: $\quad\quad\quad$ accepted ← True
13: $\quad\quad\quad$ break
14: $\quad\quad$ end if
15: $\quad\quad$ if $r < Rmax$ then
16: $\quad\quad\quad S_j^{(r+1)} \leftarrow$ LLMRevise($\tilde{S}_j^{(r)}, H_j, \Delta_j^{(r)}, \mathcal{K}_j$)
17: $\quad\quad$ end if
18: $\quad$ end for
19: $\quad$ if accepted = False then
20: $\quad\quad$ RecordScheduleFailure($H_j$, $\Delta_j^{(Rmax)}$)
21: $\quad\quad$ continue
22: $\quad$ end if
23: $\quad \mathcal{Z}_j \leftarrow$ MaterializeSchedule($S_j$, $\mathcal{E}$, $H_j$)
24: $\quad \mathcal{Z} \leftarrow \mathcal{Z} \cup \mathcal{Z}_j$; $S \leftarrow S \cup S_j$
25: end for
26: $\mathcal{M} \leftarrow$ AggregateWeeklyMetrics($\mathcal{Z}$)
27: return $\mathcal{K}, \mathcal{Z}, \mathcal{M}, S$

### 3.4 Evidence-Grounded Resident-Experience Evaluation

The Inspection and Interview module provides visual and interpretive access to the simulation results. The interface presents route maps, daily timelines, household activity histories, and summary indicators, enabling users to inspect both individual events and weekly mobility patterns.

For each selected respondent, the system filters the events in which that resident participates. It then constructs a compact evidence context containing activity summaries and representative events. Event selection is field-based rather than embedding-based, prioritizing failed trips, journeys exceeding 15 minutes, everyday service activities, and longer travel times while maintaining diversity across activity types. Each selected item can be traced directly to its underlying event record.

The interview LLM receives the question, basic respondent information, and the corresponding evidence context. It is instructed to respond in the first person and to ground its interpretation only in the supplied evidence. The backend links the response to supporting event records and identifies conditions that cannot be assessed from the available data.

**Table 4** presents a structured post-simulation questionnaire comprising seven 10-point Likert-scale items and an open-ended question on potential improvements. The questions elicit resident-specific evaluations of daily convenience, healthcare access, travel burden, activity feasibility, trip chaining, household coordination, and access to recreation and social activities. The questionnaire is not restrictive: users may design their own questions or engage in open-ended conversations with the resident agents. Readers should note that these evaluations only address the subset of neighborhood livability associated with residents' daily activities and mobility. Broader dimensions, such as the quality of social interaction, community participation, and safety, require evidence beyond that represented in the current prototype.

**Table 4.** Simulated Resident Experience Questionnaire

| Dimension | Question posed to the resident agent | Basis |
|---|---|---|
| Daily convenience | On a scale from 1 to 10, where 1 is *not at all convenient* and 10 is *extremely convenient*, how convenient was it to meet your everyday needs in this neighborhood? | Badland et al. (2014); Cerin et al. (2006) |
| Healthcare access | On a scale from 1 to 10, where 1 is *extremely difficult* and 10 is *extremely easy*, how manageable was it to access healthcare services or medicines when needed? | WHO (2007); Neutens (2015) |
| Travel burden | On a scale from 1 to 10, where 1 is *extremely burdensome* and 10 is *not at all burdensome*, how manageable were your travel times, routes, and mobility requirements? | Cerin et al. (2006); WHO (2007) |
| Activity feasibility | On a scale from 1 to 10, where 1 is *not at all feasible* and 10 is *fully feasible*, how well were you able to complete your planned activities within an acceptable amount of time? | Kwan (1999); Dong et al. (2006) |

| Trip chaining | On a scale from 1 to 10, where 1 is *not at all efficient* and 10 is *extremely efficient*, how easily could you combine necessary activities into a single outing? | Dong et al. (2006) |
|---|---|---|
| Household coordination | On a scale from 1 to 10, where 1 is *extremely difficult* and 10 is *extremely easy*, how manageable was it to coordinate work, school, shopping, and care responsibilities? | Recker et al. (2001); Neutens et al. (2008) |
| Recreation and social life | On a scale from 1 to 10, where 1 is *not at all convenient* and 10 is *extremely convenient*, how convenient was it to access parks, leisure opportunities, and social or community activities? | Badland et al. (2014); Cerin et al. (2006) |
| Potential improvements | Based on your weekly experience, what one change would most improve the livability of this neighborhood? | - |

## 4. Prototype Demonstration and Evaluation

### 4.1 Demonstration Setup

We designed a simulation experiment to demonstrate the prototype using heterogeneous resident configurations. We manually draw the boundary of Hong Leong Technology Park, a residential neighborhood in Shenzhen (**Figure 2**). Five households were instantiated from the predefined profiles in the built-in persona library (**Table 1**), and a user-defined profile representing a recently married couple without children was added as H6. **Table 5** summarizes the resulting six households and 11 resident agents. Household residences were randomly sampled within the neighborhood boundary.

**Table 5.** Household and resident agents configuration in simulation

| Household | Resident Agent | Description |
|---|---|---|
| H1: Independently living older adult | A1: Antie Chen (72,elderly) | A widowed older adult living alone. She walks slowly, undertakes mainly short daytime trips, and is sensitive to healthcare, grocery access, parks, and barrier-free mobility conditions. |
| H2: Young single professional | A2: Xiao Lin (28, adult) | A commuting young renter with fragmented weekday time availability. He relies primarily on walking and public transport for everyday services, shopping, and recreational activities. |
| H3: Dual-earner household with infant | A3: Ming Zhou (34, adult)<br>A4: Qian Zhou (32, adult)<br>A5: Xiaobao Zhou (1, infant) | A dual-earner household with an infant. Caregivers face longer and more constrained activity patterns, with particular needs related to childcare and health services, grocery shopping, parks, and stroller-friendly travel. |
| H4: Household with school-age child | A6: Qiang Wu (38, adult)<br>A7: Fang Wu (36, adult)<br>A8: You Wu (9, child) | A couple with a primary-school child. School escort, after-school activities, and grocery shopping create coordinated daily travel demands, allowing examination of the |

|  |  | relationship between school access and local community services. |
|---|---|---|
| H5: Mobility-constrained adult | A9: Mr. Gao (58, adult) | An adult with lower-limb mobility limitations who depends on slow walking and nearby services. This profile is used to examine barrier-free mobility and access to local healthcare and grocery facilities. |
| H6: Married young couple without children (user-customized) | A10: Xiao Hong (27, adult)<br>A11: Xiao Ming (29, adult) | A recently married young couple without children. This user-defined household represents working adults with evening social, leisure, and everyday shopping needs. |

**Figure 2** shows the prototype's web interface. The upper-left pane provides an interactive map for defining the neighborhood boundary and visualizing roads, facilities, and simulated activity-mobility trajectories, with a timebar showing resident location over time. The lower-left pane summarizes simulation outputs, including trip counts, completed and failed activities, the share of trips completed within 15 minutes, and mean travel time; it also allows users to filter the event timeline by day. The central pane displays the built-in and custom persona library for users to add and select households in simulation. The right-hand pane is a chatroom, where users can ask structured or open-ended questions and review responses from selected resident agents. Users can choose to save the simulation results in JSON file after simulation completion; the file can be reloaded and queried later in the interface.

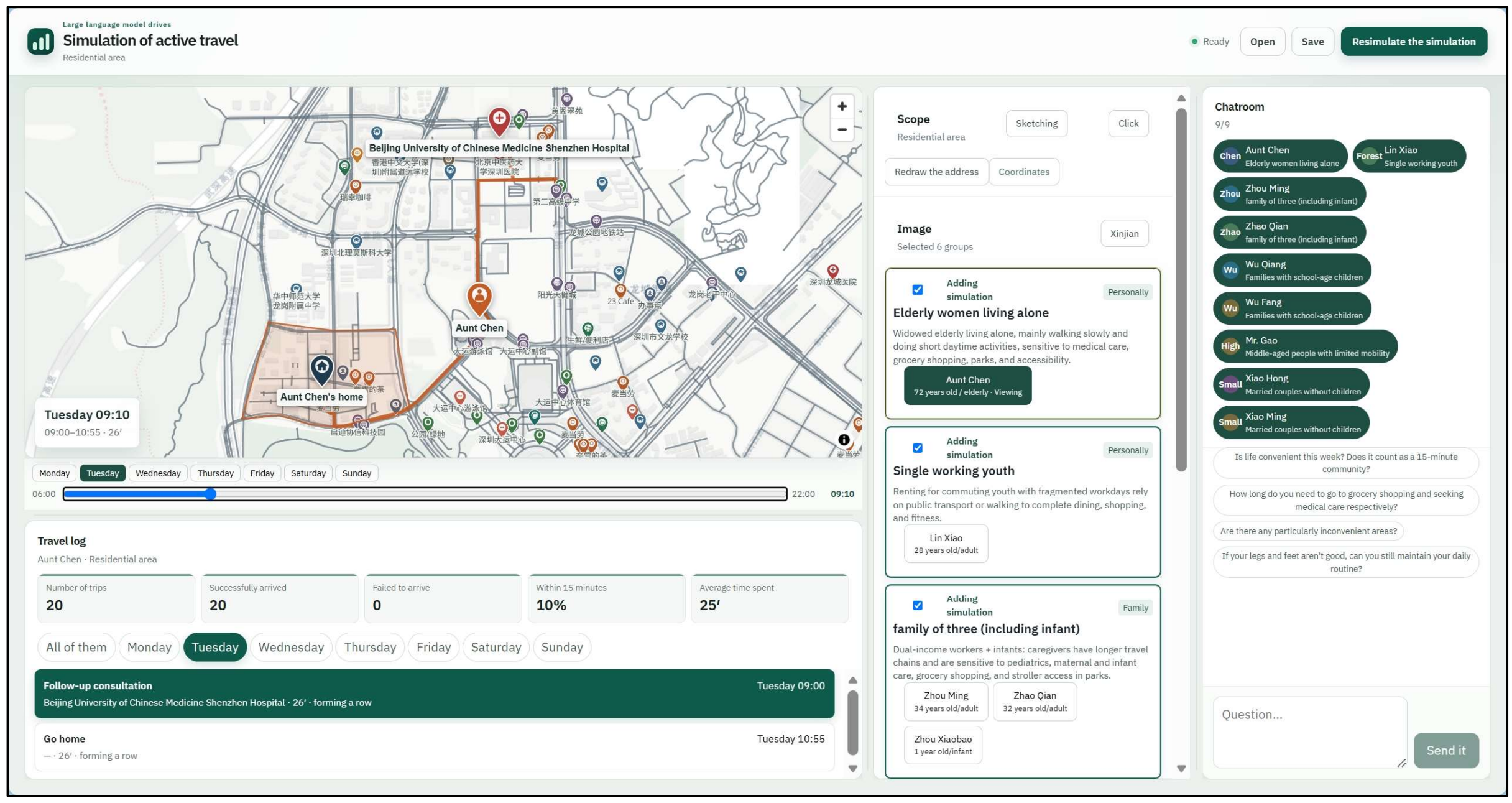

**Figure 2.** Prototype interface (translated in English for presentation).

## 4.2 Schedule and event outcomes

We examine the weekly simulation trace of Aunt Chen (A1), an independently living older woman. Her schedule contains 12 out-of-home activities and eight return-home events, yielding 20 activity–mobility records over the simulation week. The activities include grocery shopping, park visits, community activities, healthcare and pharmacy visits, and banking. All scheduled activities were completed. However, her mean travel time was 25.1 minutes, and only 10.0% of her trips fell within the 15-minute threshold. This result shows that Aunt Chen could reach all required destinations, but doing so required repeated trips throughout the week. Specifically, her activity-specific mean travel times ranged from 9.8 minutes for pharmacy visits to 31.3 minutes for park visits. The neighborhood was therefore serviceable, but potentially burdensome for a resident with reduced mobility. **Figure 3** visualizes the weekly activity-mobility log for Aunt Chen (A1). The activity logs for other resident agents can be found in the **Appendix A**.

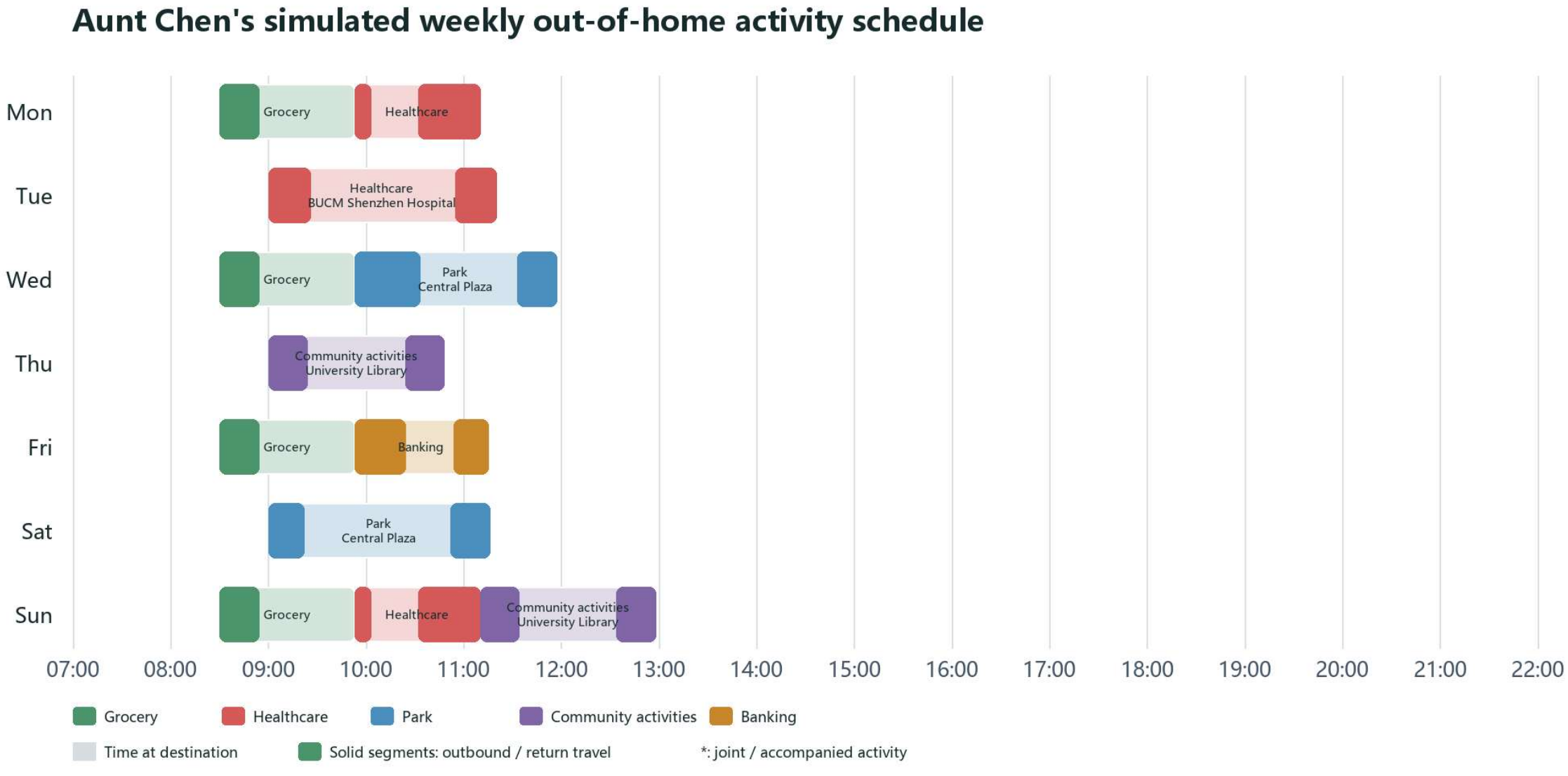

**Figure 3**. Weekly activity-mobility log for resident agent Aunt Chen (A1).

**Table 6** reports resident-level outcomes for all agents. The counts include both out-of-home activities and return-home events. Joint events are attributed to every participating resident. Across the 11 agents, the simulation generated 172 unique event records.

**Table 6.** Resident-level schedule and event outcome.

| Profile | Planned trips | Completed trips | Failed/substituted activities | Mean travel time (min) | Trips within 15 min (%) | Accompanied trips |
|---|---|---|---|---|---|---|
| A1: Aunt Chen | 20 | 20 | 0 | 25.1 | 10.0 | 0 |
| A2: Xiao Lin | 23 | 23 | 0 | 20.4 | 21.7 | 0 |
| A3: Ming Zhou | 27 | 27 | 0 | 22.7 | 3.7 | 22 |
| A4: Qian Zhou | 23 | 23 | 0 | 18.5 | 30.4 | 15 |
| A5: Xiaobao Zhou | 22 | 22 | 0 | 23.2 | 4.5 | 22 |

| A6: Qiang Wu | 32 | 32 | 0 | 30.7 | 21.9 | 27 |
|---|---|---|---|---|---|---|
| A7: Fang Wu | 23 | 23 | 0 | 20.1 | 30.4 | 16 |
| A8: You Wu | 27 | 27 | 0 | 25.6 | 25.9 | 27 |
| A9: Mr. Gao | 22 | 22 | 0 | 22.6 | 36.4 | 0 |
| A10: Xiao Hong | 22 | 22 | 0 | 17.9 | 36.4 | 4 |
| A11: Xiao Ming | 15 | 15 | 0 | 21.5 | 33.3 | 4 |

Travel burdens varied substantially across agents. Qiang Wu (A6) had the highest mean travel time (30.7 minutes), whereas Xiao Hong (A10) had the lowest (17.9 minutes). Aunt Chen recorded only 10.0% of trips within 15 minutes, compared with 36.4% for Mr. Gao (A9) and Xiao Hong (A10). These results show that travel burden is shaped jointly by persona attributes, activity schedules, household responsibilities, and spatial location. Because home locations were randomly sampled within the neighborhood boundary, the observed differences also reflect residential micro-location.

Household outcomes reveal that accessibility is partly dependent on household coordination. All events involving the infant (A5) and school-age child (A8) were accompanied. Qiang Wu (A6) participated in 27 accompanied events, while Fang Wu (A7) participated in 16. These records include school escort, shopping, leisure, and return-home travel. Accessibility for children and care-dependent household members therefore depends not only on nearby facilities, but also on the availability of another household member.

No activity failed or required substitution in this simulation. This indicates that all scheduled needs had routable facility options in the selected environment, but it does not establish that the neighborhood is universally livable. The current scenario is more informative about differences in travel burden and household coordination than about severe service deprivation.

### 4.3 Evidence-grounding audit for resident agents' living experiences

Following the simulation, we surveyed the nine adult resident agents regarding their living experiences in the simulated week using the questions in **Table 4**. **Figure 4** shows the interactive interface, through which users can add one or more resident agents to the chatroom for interviewing.

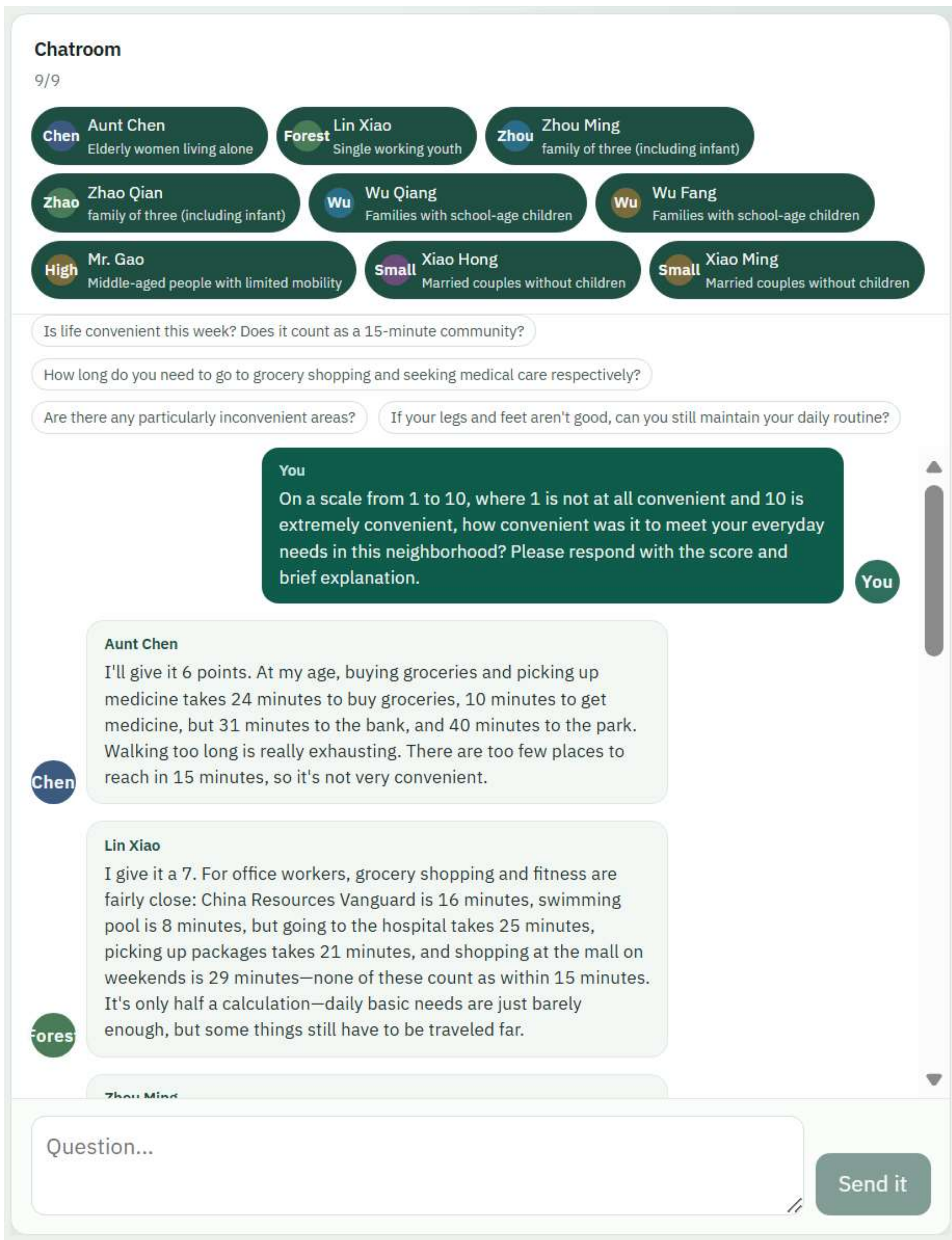

**Figure 4**. Chatroom interface for querying simulated resident agents (The interface is translated in English for presentation).

**Table 7** reports the agents' ratings of the different aspects of neighborhood livability. These scores should not be interpreted as survey observations or validated measures of real perceptions. Instead, they provide a structured summary of how different simulated residents interpreted their own event histories. A dash indicates that the relevant experience was not represented in an agent's simulated event record. Based on **Table 7**, activity feasibility and travel burden received the highest mean scores (both 6.4), whereas healthcare access received the lowest (5.0). This lower rating reflects the absence of a major public hospital near the residential area, requiring agents to travel more than two kilometers to the nearest hospital. Some pharmacies were located close to grocery stores and could therefore be incorporated into chained outing. Trip chaining also received a relatively low rating (5.2), as many destinations were spatially dispersed in different directions. Daily convenience, household coordination, and recreation received intermediate mean ratings of 6.0-6.2. Overall, the agents were generally able to complete their scheduled activities, but the spatial distribution of facilities continued to constrain healthcare access and efficient activity chaining.

Ratings also varied across resident profiles. Mr. Gao (A9) provided the lowest ratings across most dimensions, particularly healthcare access (2) and trip chaining (3), consistent with his mobility constraints. Ming Zhou (A3) also reported relatively low ratings, reflecting the travel and coordination demands of a household with an infant. Aunt Chen rated healthcare access below the sample mean, although her daily-convenience rating equaled the mean. In contrast, Xiao Hong (A10) reported consistently high ratings, particularly for activity feasibility, household coordination, and recreation. Trip chaining remained comparatively weak across the sample, with scores ranging from 3 to 7.

**Table 7.** Resident-agent evaluations of neighborhood livability

| Profile | Daily convenience | Healthcare access | Travel burden | Activity feasibility | Trip chaining | Household coordination | Recreation and social life |
|---|---|---|---|---|---|---|---|
| A1: Aunt Chen | 6 | 4 | 6 | 6 | 5 | - | 6 |
| A2: Xiao Lin | 7 | 6 | 7 | 7 | 6 | - | 7 |
| A3: Ming Zhou | 5 | 3 | 5 | 5 | 4 | 4 | 5 |
| A4: Qian Zhou | 6 | 4 | 6 | 6 | 5 | 5 | 6 |
| A6: Qiang Wu | 6 | 5 | 7 | 7 | 6 | 6 | 6 |
| A7: Fang Wu | 6 | 6 | 7 | 8 | 5 | 7 | 6 |
| A9: Mr. Gao | 4 | 2 | 4 | 4 | 3 | - | 4 |
| A10: Xiao Hong | 7 | 7 | 8 | 8 | 7 | 8 | 8 |
| A11: Xiao Ming | 7 | 8 | 8 | 7 | 6 | 7 | 7 |
| Average | 6.0 | 5.0 | 6.4 | 6.4 | 5.2 | 6.2 | 6.1 |

**Table 8** summarizes the agents' open-ended suggestions for neighborhood improvement and provides qualitative context for the ratings in **Table 7**. The responses reveal resident-specific priorities: older and mobility-constrained agents emphasized closer healthcare, parks, and accessible walking conditions; households with children highlighted childcare-supportive facilities, nearby recreation, and convenient daily services; and younger adults more frequently requested parcel lockers, community services, and leisure or social destinations. These responses show how the interview component translates activity–mobility records into differentiated planning concerns.

These responses remain conditional on the simulation inputs and should be interpreted cautiously. The experiment represents only one simulated week and relies on OSM-downloaded facility and network data, which may contain classification errors or omit facilities or access restrictions. For example, Yudiefang was classified as a grocery destination although it is a bakery, while a frequently selected 7-Eleven is located within a university campus and may not be publicly accessible. Also, most residential areas have parcel collection points within the community in China, which was not marked in OSM data. More accurate facility data, home-location assignments, and scheduling assumptions could alter both the simulated event records and subsequent evaluations. Therefore, the interview outputs represent traceable interpretations of a modeled environment rather than direct evidence of residents' perceptions or policy preferences.

**Table 8.** Resident-agent suggestion on community improvement

| | **Response** |
|---|---|
| A1: Aunt Chen | What I hope most is to have a small vegetable shop or a community canteen right at my doorstep. This week, it took me 24 minutes to buy groceries and walk to Yudie Fang, and although it took 10 minutes to get medicine to Neptune, I still had to cross the street. My legs and feet really couldn't handle it after walking for so long. If I could buy groceries and get some common medicine downstairs, I wouldn't have to travel so far. |
| A2: Xiao Lin | I suggest moving the delivery points closer or setting up more pickup lockers. This week, I went to the post office to pick up a package and walked for 21 minutes—almost an hour round trip. I had to rush back after work, which wasted too much time. If I could pick up the package in the neighborhood, I'd just take a stroll at night and pick it up on the way. |
| A3: Ming Zhou | Most importantly, pediatric outpatient clinics or community clinics are preferred. On Sunday, the child had a fever and went to Beijing University of Chinese Medicine Shenzhen Hospital, took a 23-minute bus ride, squeezing onto a bus with the child in their arms—truly a pain. If there were a clinic in the community that could treat minor illnesses, we wouldn't have to travel so far. |
| A4: Qian Zhou | I agree with Zhou Ming—taking a child to see a doctor is too exhausting. Also, on Saturday, I spent 21 minutes shopping at Dayun World. Although it wasn't too far, pushing a stroller was no easy task. If there were a mother and baby products store or a small supermarket in the community, daily shopping would be much more convenient. |
| A6: Qiang Wu | I want to bring the park closer. On Saturday, I took my family for a 28-minute walk in the central square, and the kids were exhausted. If there were a small park or green space next to the community, we could take a stroll after dinner without having to walk so far. |
| A7: Fang Wu | I also felt the park was too far. On Saturday, I walked 28 minutes to the central square, and with my child, it took almost an hour to go back and forth. If there were a small garden or fitness corner in the community, we could take our kids out to get around every day, without waiting until weekends. |
| A9: Mr. Gao | What I need most is to move the pharmacy closer. I went to Neptune twice this week to get medicine, each time walking for 33 minutes. My legs are inconvenient, and the distance is really exhausting. If there were a pharmacy or health station in |
| | the community, I wouldn't have to go through all this trouble to get medicine. |
| A10: Xiao Hong | I wish the barbershop could be a bit closer. On Thursday, I went to the post office for a haircut and walked for 22 minutes, so the round trip took almost fifty minutes. If there were a barbershop in the neighborhood or downstairs, I could cut my hair on the way home from work without having to make a special trip. |
| A11: Xiao Ming | I suggest moving the post office or delivery point closer together. This week, I went to the post office to pick up a package and walked for 31 minutes, taking over an hour round trip. It was such a waste of time. If there were a collection point in the community, I could pick it up on my way after work—much more convenient. |

## 5. Discussion

This study developed and demonstrated an LLM-agent prototype that combines a spatial KG, structured household personas, LLM-based schedule generation and revision, rule-supported feasibility review, network materialization, and evidence-grounded interviews to assess neighborhood livability through simulated resident experience.

Unlike conventional agent-based models (e.g., Hao et al., 2022) that typically rely on predefined behavioral rules or estimated choice models, the present prototype uses LLMs to generate household schedules from structured personas and retrieved spatial context. The deterministic rule layer does not replace behavioral modeling; rather, it constrains and audits the LLM outputs. The framework should therefore be understood as a hybrid generative-symbolic prototype rather than a calibrated travel-demand model. Building on this design, the prototype extends conventional GIS-based livability assessment by treating accessibility as an activity-based and household-dependent process rather than a static property of a residential location. Existing measures of proximity, reachable opportunities, and threshold-based access remain useful for describing neighborhood opportunity structures (Geurs & van Wee, 2004). However, everyday travel is rarely composed of independent home-to-facility trips. Residents coordinate activities with work and school commitments, accompany children or dependents, and undertake linked activities across time and space. Activity-based and joint-accessibility research has long emphasized these temporal and interpersonal constraints (Dong et al., 2006; Kwan, 1999; Neutens et al., 2008), but they are seldom represented in static neighborhood-livability assessments. By retrieving a household-centered ego-subgraph before schedule generation, the prototype allows local destination availability and approximate travel burdens to inform activity timing, sequencing, and chaining. Spatial context therefore enters the planning process rather than being applied only after an abstract schedule has been generated.

The simulation results illustrate this gap. All scheduled activities were completed in the presented scenario, yet the resulting experience was not equally convenient for all agents. Aunt Chen and Mr. Gao, the mobility-sensitive profiles, had no trips within 15 minutes though reaching their required destinations. Household outcomes further showed that all simulated trips involving the infant and school-age child required accompaniment. These patterns illustrate that facility availability is not equivalent to usable access: the practical burden of access depends on travel time, mobility capacity, activity timing, and the availability of other household members. The prototype therefore provides a way to examine how the same spatial environment produces different mobility burdens and coordination demands across resident profiles.

The resident-interview component adds a second contribution. Rather than allowing an LLM to generate unrestricted accounts of neighborhood life, the prototype links responses to structured simulation events. This design follows the broader move from purely generative agents toward agents constrained by explicit environments, retrieved context, verification mechanisms (Park et al., 2023). In the present study, the distinction is especially important because livability is partly subjective: agents can interpret simulated activities as convenient or burdensome, but these interpretations remain linked to recorded routes, travel times, destinations, outcomes, and companionship. The interviews should not be read as substitutes for resident surveys; instead, they provide an inspectable intermediate layer between event-based simulation output and resident-specific interpretation.

More broadly, the framework suggests a direction for future urban digital-twin and planning applications. Static GIS indicators can identify uneven opportunity structures, while spatially grounded LLM agents can reveal the temporal and household consequences of those structures. Evidence-grounded LLM interviews may then help planners interpret those consequences in resident-centred terms. This combination could support comparative analyses of multiple neighborhoods, scenario testing for facility or transport interventions, and the identification of resident groups for whom nominal proximity does not translate into manageable daily life.

Several limitations should guide future work. First, simulation outcomes depend on the accuracy of OSM-derived roads and facilities. Randomly sampled home locations can also alter routes, schedules, and interview responses. Second, the model has not been calibrated against observed trajectories. Mobility and stay data could improve group-specific travel preferences, departure times, destination choices, and commonly linked activity chains. Third, schedule generation and interview responses remain sensitive to the selected LLM, prompts, and review settings. Repeated runs and sensitivity analysis are therefore needed to assess output stability. Finally, the current assessment focuses on mobility, facilities, and accessibility. Livability also includes social interaction, safety, affordability, environmental quality, and other experiential dimensions. Introducing interaction mechanisms between agents and additional contextual data would allow future versions to represent social encounters and community participation.

## 6. Conclusion

This study presents a spatial-KG-grounded LLM-agent prototype for evaluating neighborhood livability through simulated resident experience. The framework combines household-level LLM schedule generation and revision, rule-based feasibility review, network materialization, and evidence-grounded resident interviews. Its main contribution is to connect structured spatial context with personalized activity-mobility simulation and traceable resident-specific evaluation, thereby extending neighborhood assessment beyond static facility proximity to consider how mobility capacities, schedules, activity chains, and household responsibilities shape usable access. The prototype demonstration reveals when nominally accessible opportunities impose unequal travel or coordination burdens across resident groups. The prototype can support neighborhood comparison, intervention testing, and the identification of residents whose daily needs are poorly served by existing spatial configurations. Future research will calibrate the prototype using observed mobility and survey data, evaluate its behavioral validity, and apply standardized personas across multiple neighborhoods.

**References**

Abbiasov, T., Heine, C., Sabouri, S., Salazar-Miranda, A., Santi, P., Glaeser, E., & Ratti, C. (2024). The 15-minute city quantified using human mobility data. *Nature Human Behaviour*, *8*(3), 445-455.

Badland, H., & Pearce, J. (2019). Liveable for whom? Prospects of urban liveability to address health inequities. *Social science & medicine*, *232*, 94-105.

Badland, H., Whitzman, C., Lowe, M., Davern, M., Aye, L., Butterworth, I., Hes, D., & Giles-Corti, B. (2014). Urban liveability: Emerging lessons from Australia for exploring the potential for indicators to measure the social determinants of health. *Social Science & Medicine, 111*, 64–73. https://doi.org/10.1016/j.socscimed.2014.04.003

Cerin, E., Saelens, B. E., Sallis, J. F., & Frank, L. D. (2006). Neighborhood Environment Walkability Scale: Validity and development of a short form. *Medicine & Science in Sports & Exercise, 38*(9), 1682–1691. https://doi.org/10.1249/01.mss.0000227639.83607.4d

Dong, X., Ben-Akiva, M. E., Bowman, J. L., & Walker, J. L. (2006). Moving from trip-based to activity-based measures of accessibility. *Transportation Research Part A: Policy and Practice, 40*(2), 163–180. https://doi.org/10.1016/j.tra.2005.05.002

Edge, D., Trinh, H., Cheng, N., Bradley, J., Chao, A., Mody, A., ... & Larson, J. (2024). From local to global: A graph rag approach to query-focused summarization. *arXiv preprint arXiv:2404.16130*.

Geurs, K. T., & Van Wee, B. (2004). Accessibility evaluation of land-use and transport strategies: review and research directions. *Journal of Transport geography*, *12*(2), 127-140.

Hao, H., & Wang, Y. (2022). Disentangling relations between urban form and urban accessibility for resilience to extreme weather and climate events. *Landscape and urban planning*, *220*, 104352.

Hao, H., Wang, Y., & Chen, J. (2024). Empowering scenario planning with artificial intelligence: A perspective on building smart and resilient cities. *Engineering*, *43*, 272-283.

Hao, H., Wang, Y., & Wang, Q. (2022). Simulating urban population activities under extreme events with data-driven agent-based modeling. In *Construction Research Congress 2022* (pp. 1125-1134).

Khavarian-Garmsir, A. R., Sharifi, A., & Sadeghi, A. (2023). The 15-minute city: Urban

planning and design efforts toward creating sustainable neighborhoods. *Cities*, *132*, 104101.

Kwan, M.-P. (1998). Space-time and integral measures of individual accessibility: A comparative analysis using a point-based framework. *Geographical Analysis, 30*(3), 191–216. https://doi.org/10.1111/j.1538-4632.1998.tb00396.x

Kwan, M.-P. (1999). Gender and individual access to urban opportunities: A study using space–time measures. *The Professional Geographer, 51*(2), 210–227. https://doi.org/10.1111/0033-0124.00158

Lämmer, S., Colley, M., & Ebel, P. (2026, April). GTA: Generative traffic agents for simulating realistic mobility behavior. In *Proceedings of the 2026 CHI Conference on Human Factors in Computing Systems* (pp. 1-16).

Liu, Q., Li, C., & Ma, W. (2026). GATSim: Urban mobility simulation with generative agents. *Transportation Research Part C: Emerging Technologies*, *186*, 105576.

Logan, T. M., Hobbs, M. H., Conrow, L. C., Reid, N. L., Young, R. A., & Anderson, M. J. (2022). The x-minute city: Measuring the 10, 15, 20-minute city and an evaluation of its use for sustainable urban design. *Cities*, *131*, 103924.

McMillan, T. E. (2005). Urban form and a child's trip to school: The current literature and a framework for future research. *Journal of Planning Literature, 19*(4), 440–456. https://doi.org/10.1177/0885412204274173

Moreno, C., Allam, Z., Chabaud, D., Gall, C., & Pratlong, F. (2021). Introducing the 15-minute city: Sustainability, resilience and place identity in future post-pandemic cities. *Smart Cities, 4*(1), 93–111. https://doi.org/10.3390/smartcities4010006

Neutens, T. (2015). Accessibility, equity and health care: Review and research directions for transport geographers. *Journal of Transport Geography, 43*, 14–27.

Neutens, T., Schwanen, T., Witlox, F., & De Maeyer, P. (2008). My space or your space? Towards a measure of joint accessibility. *Computers, Environment and Urban Systems, 32*(5), 331–342. https://doi.org/10.1016/j.compenvurbsys.2008.06.001

Park, J. S., O'Brien, J., Cai, C. J., Morris, M. R., Liang, P., & Bernstein, M. S. (2023, October). Generative agents: Interactive simulacra of human behavior. In *Proceedings of the 36th annual acm symposium on user interface software and technology* (pp. 1-22). https://doi.org/10.1145/3586183.3606763

Patwary, A. U. Z., Ciari, F., Angioloni, L., Naseri, H., Brusci, L., & Iannelli, G. (2026). Bridging AI and Traffic Simulation: A Robust and Comprehensive Framework for LLM-Based AI Replanning Agents in MATSim. *Procedia Computer Science*, *280*, 622-629.

Pozoukidou, G., & Chatziyiannaki, Z. (2021). 15-Minute City: Decomposing the new urban planning eutopia. *Sustainability, 13*(2), 928. https://doi.org/10.3390/su13020928

Recker, W. W., Chen, C., & McNally, M. G. (2001). Measuring the impact of efficient household travel decisions on potential travel time savings and accessibility gains. *Transportation Research Part A: Policy and Practice*, *35*(4), 339-369.

Santos, G. H., Viana, A. C., & Silva, T. H. (2026). *When plausible is not realistic: Evaluating human mobility in LLM-based urban simulation* [Preprint]. arXiv. https://arxiv.org/abs/2606.13835

Song, Y.-L., Tsern, C.-E., Wu, C.-C., Chang, Y.-M., Huang, S.-B., Lin, W.-C., Lin, Y.-T., & Lin, M. C.-L. (2025). *Incorporating LLMs for large-scale urban complex mobility simulation* [Preprint]. arXiv. https://arxiv.org/abs/2505.21880

Wang, H., Tsoi, K. H., & Loo, B. P. (2025). An assessment framework for 15-minute Cities: Progress worldwide and the impact of urban form. *Transportation Research Part A: Policy and Practice*, *199*, 104583.

World Health Organization. (2007). *Global age-friendly cities: A guide*.

Wu, S., Chen, B., An, J., Nelson, A., Dai, F., Lin, C., & Gong, P. (2025). Measuring global human accessibility to essential daily necessities and services. *Nature communications*, *16*(1), 10709.

Zhang, Y., & Kwan, M. P. (2025). Considering human mobility is essential for accurate environmental exposure assessment. *Science bulletin*.

Zhou, Z., Lin, Y., Jin, D., & Li, Y. (2024). *Large language model for participatory urban planning* [Preprint]. arXiv. https://arxiv.org/abs/2402.17161

# Appendix A

## Xiao Lin's simulated weekly out-of-home activity schedule

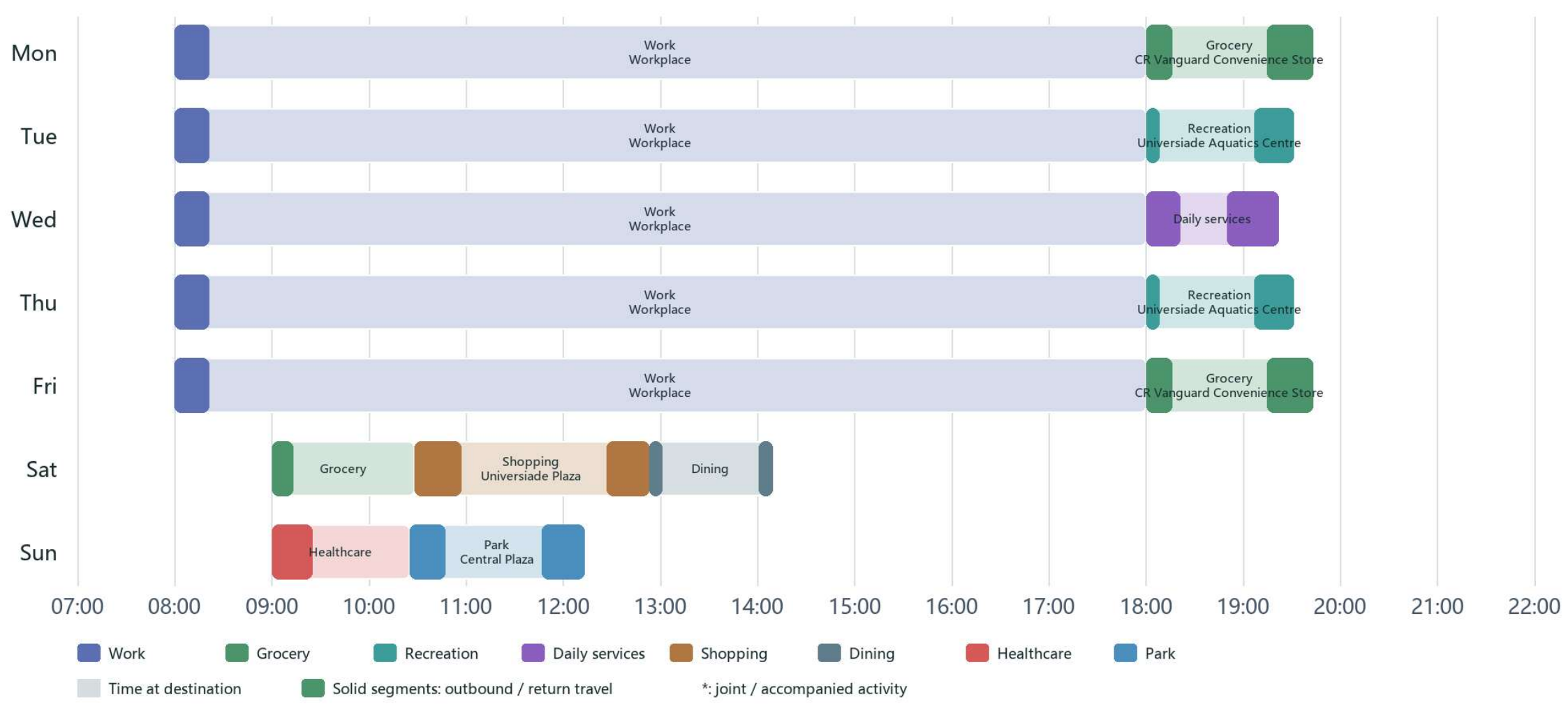

## Ming Zhou's simulated weekly out-of-home activity schedule

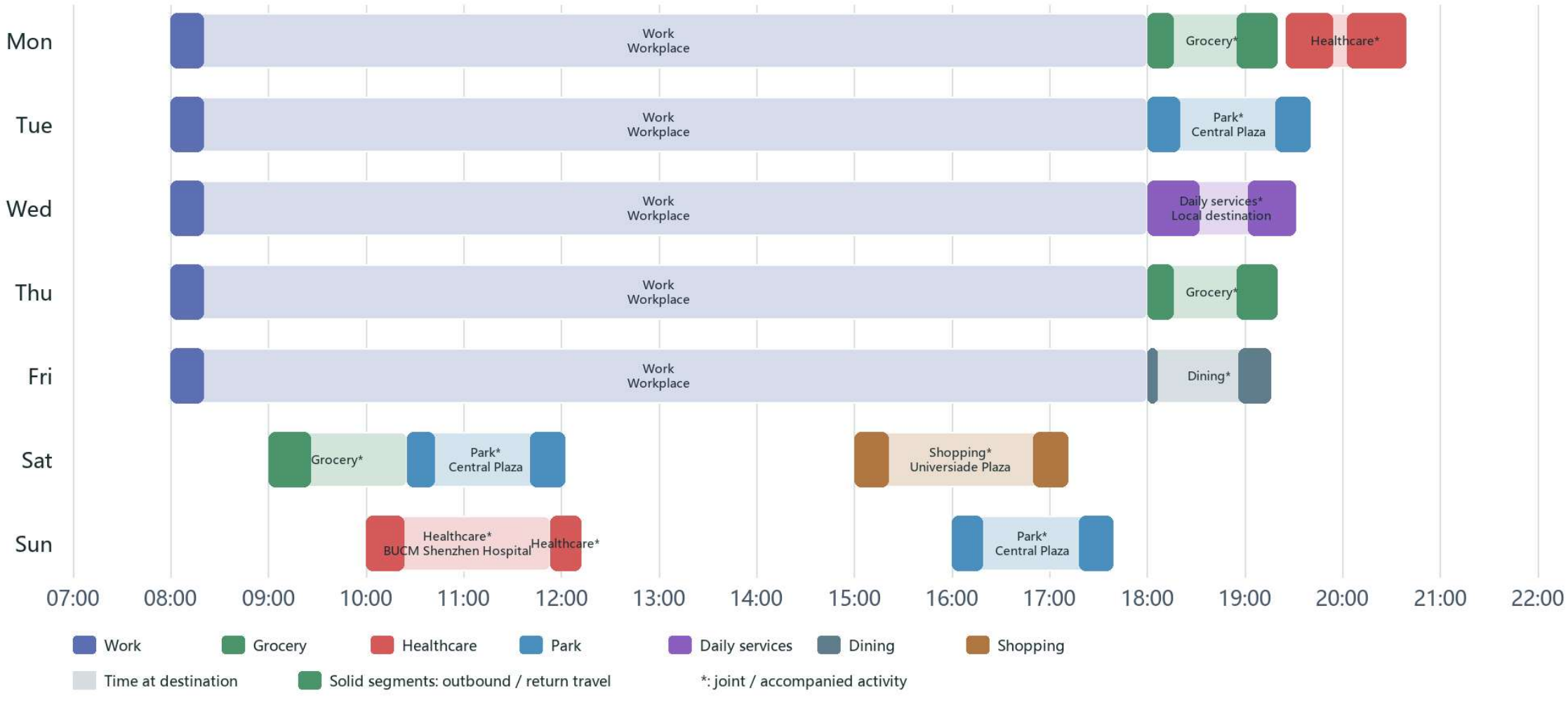

## Qian Zhou's simulated weekly out-of-home activity schedule

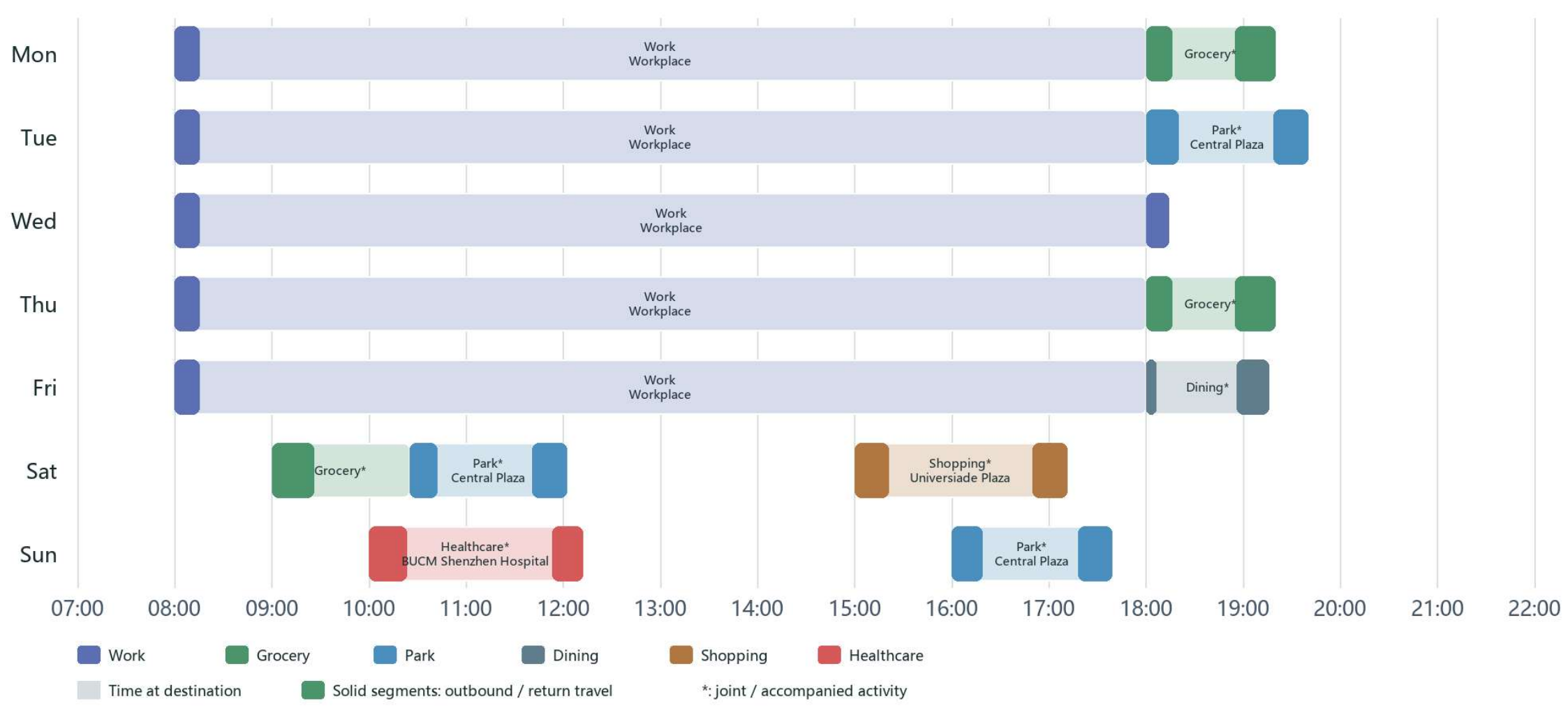

## Xiaobao Zhou's simulated weekly out-of-home activity schedule

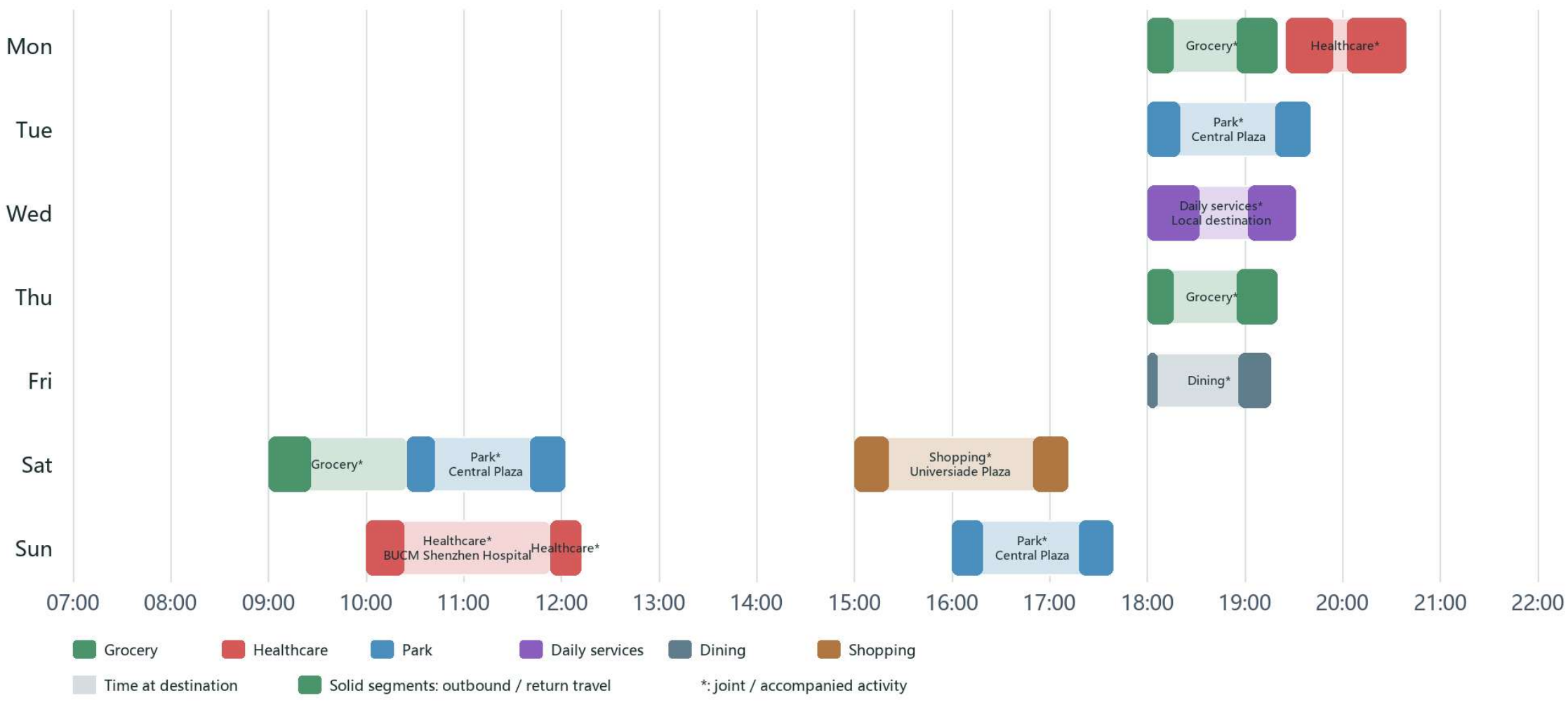

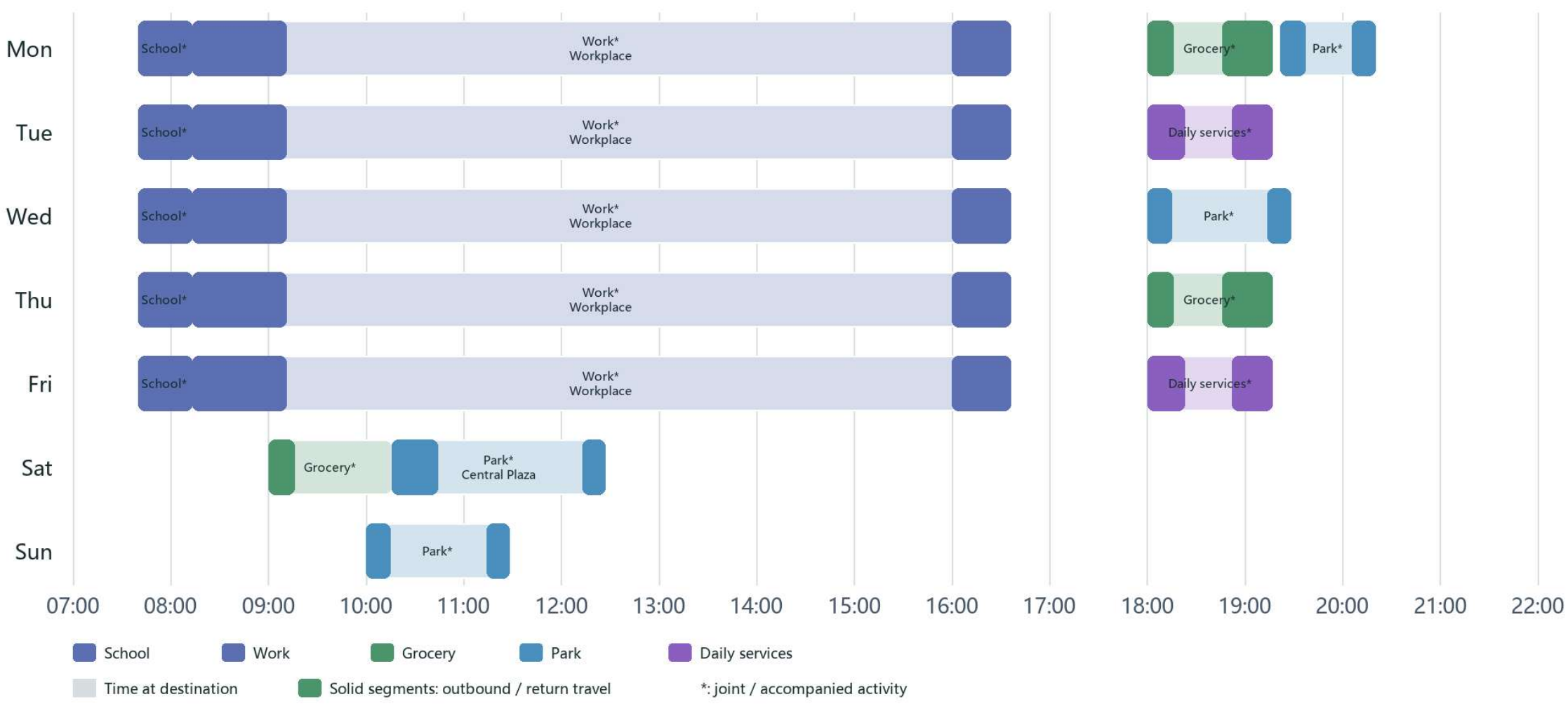
Qiang Wu's simulated weekly out-of-home activity schedule
Mon
Tue
Wed
Thu
Fri
Sat
Sun
School*
Work*
Workplace
Grocery*
Park*
Daily services*
Park*
Central Plaza
07:00
08:00
09:00
10:00
11:00
12:00
13:00
14:00
15:00
16:00
17:00
18:00
19:00
20:00
21:00
22:00
School
Work
Grocery
Park
Daily services
Time at destination
Solid segments: outbound / return travel
*: joint / accompanied activity

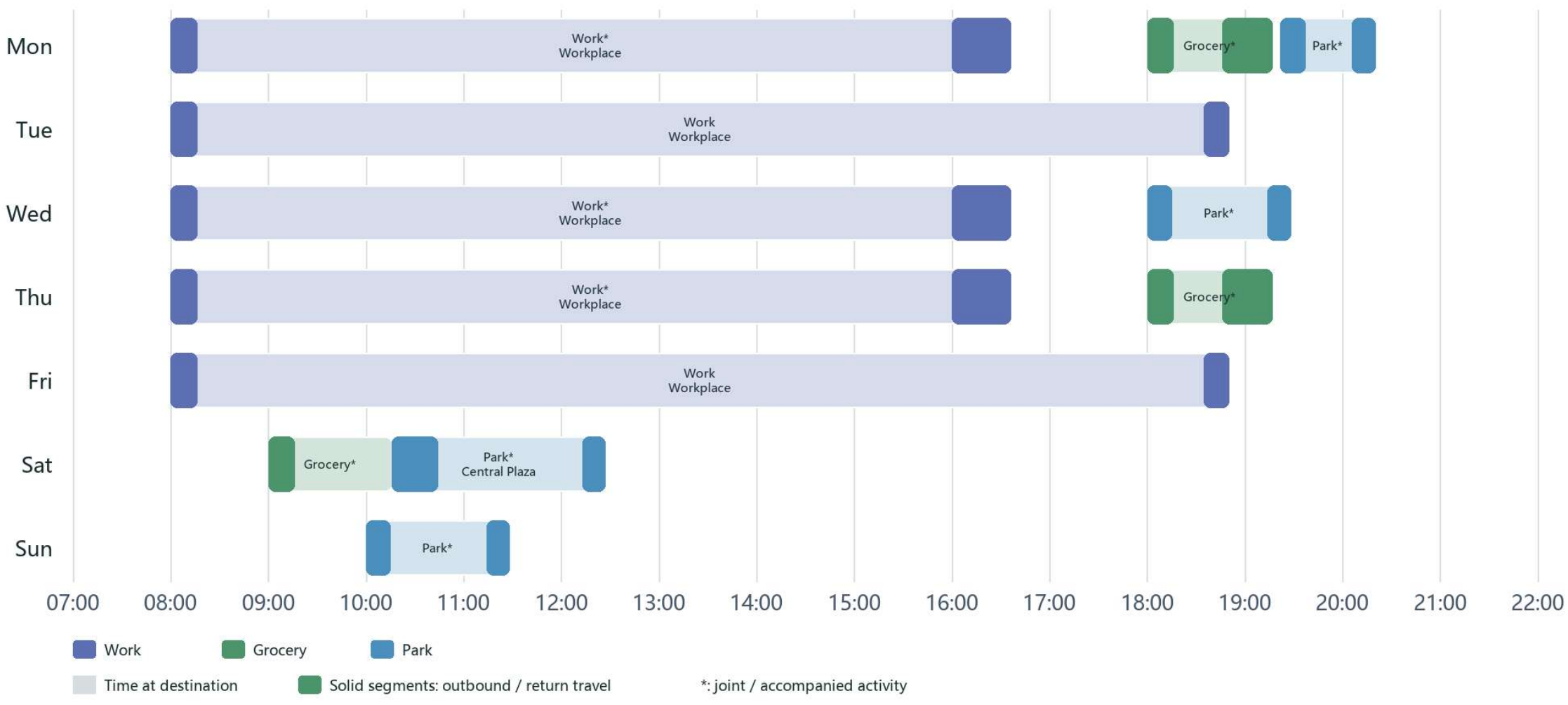
Fang Wu's simulated weekly out-of-home activity schedule
Mon
Tue
Wed
Thu
Fri
Sat
Sun
Work*
Workplace
Work
Workplace
Grocery*
Park*
Park*
Central Plaza
07:00
08:00
09:00
10:00
11:00
12:00
13:00
14:00
15:00
16:00
17:00
18:00
19:00
20:00
21:00
22:00
Work
Grocery
Park
Time at destination
Solid segments: outbound / return travel
*: joint / accompanied activity

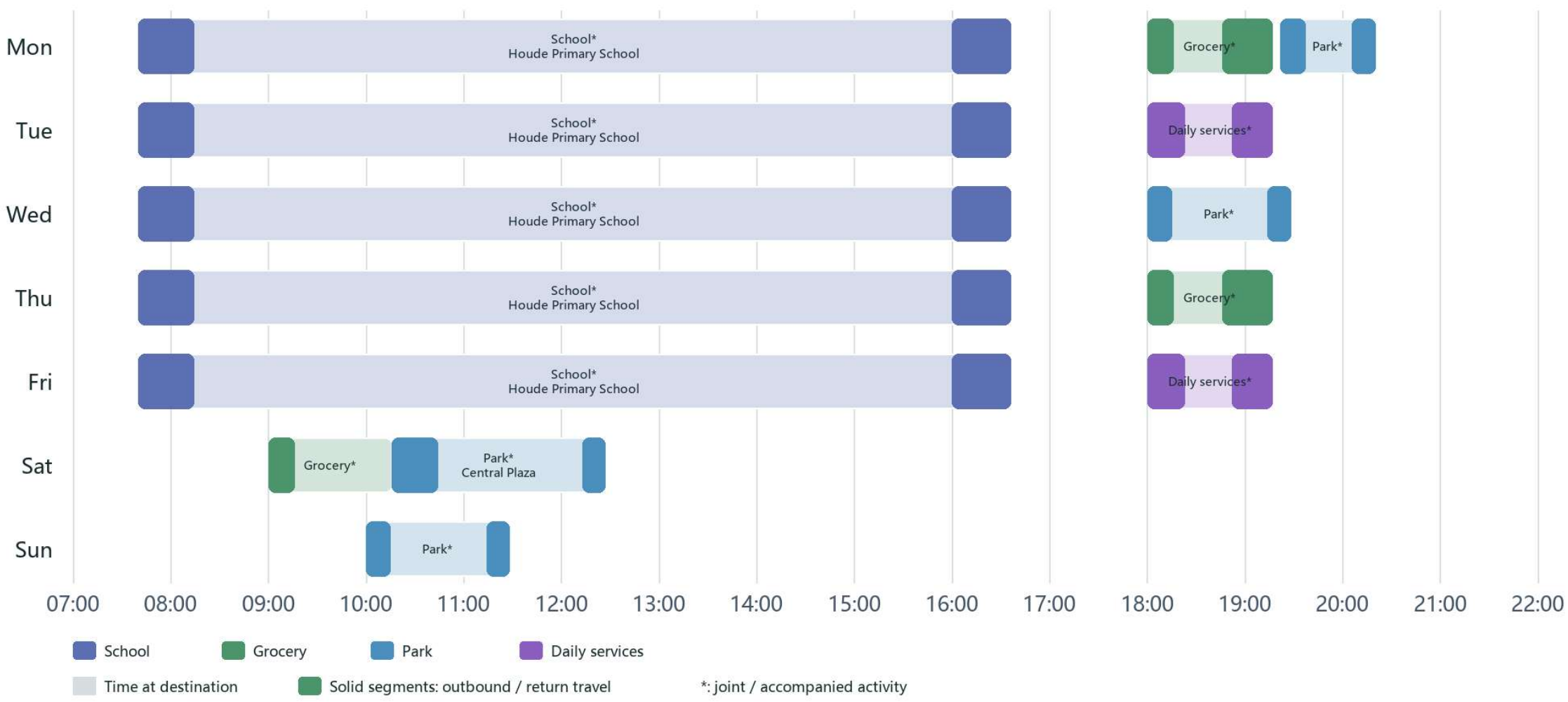
You Wu's simulated weekly out-of-home activity schedule
Mon
Tue
Wed
Thu
Fri
Sat
Sun
School*
Houde Primary School
Grocery*
Park*
Daily services*
Central Plaza
07:00
08:00
09:00
10:00
11:00
12:00
13:00
14:00
15:00
16:00
17:00
18:00
19:00
20:00
21:00
22:00
School
Grocery
Park
Daily services
Time at destination
Solid segments: outbound / return travel
*: joint / accompanied activity

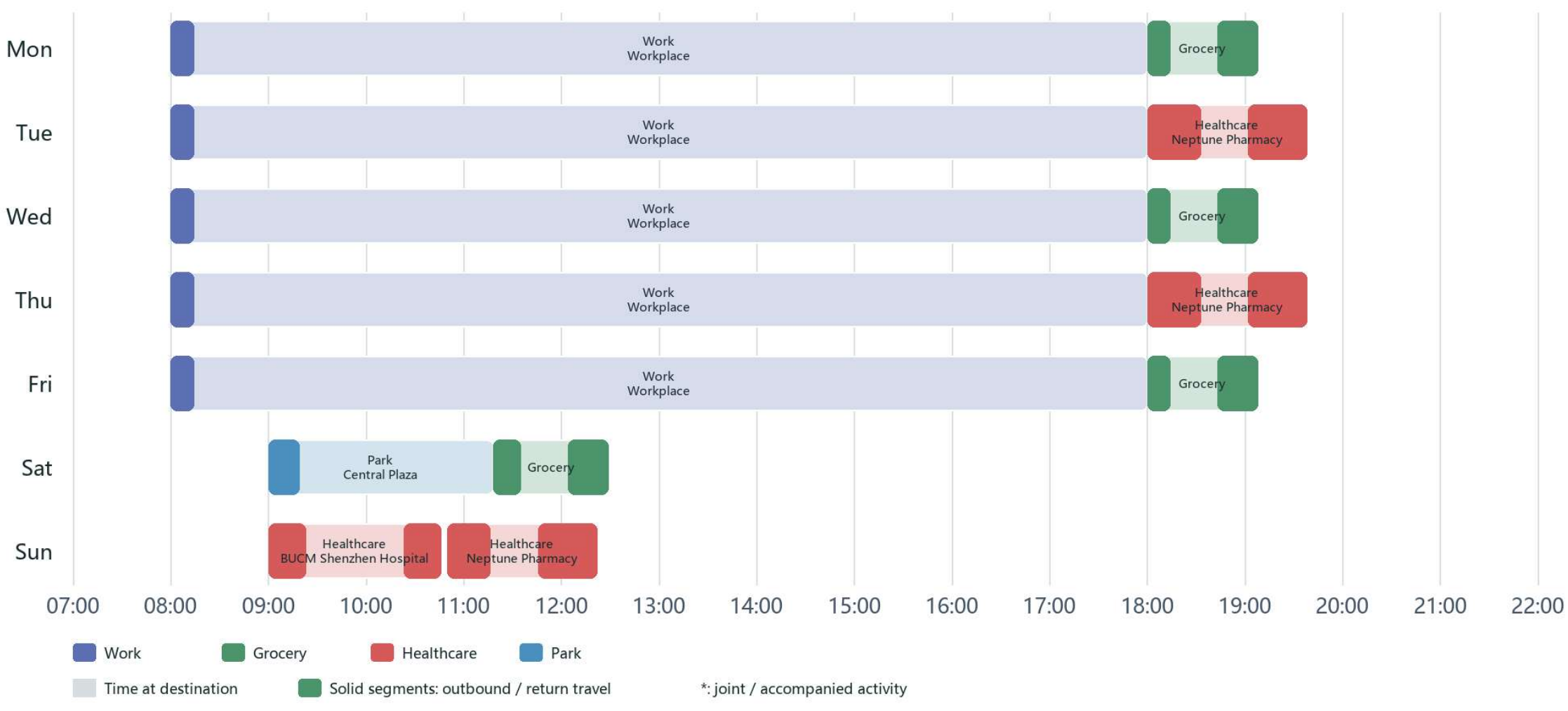
Mr. Gao's simulated weekly out-of-home activity schedule
Mon
Tue
Wed
Thu
Fri
Sat
Sun
Work
Workplace
Grocery
Healthcare
Neptune Pharmacy
Park
Central Plaza
BUCM Shenzhen Hospital
07:00
08:00
09:00
10:00
11:00
12:00
13:00
14:00
15:00
16:00
17:00
18:00
19:00
20:00
21:00
22:00
Work
Grocery
Healthcare
Park
Time at destination
Solid segments: outbound / return travel
*: joint / accompanied activity

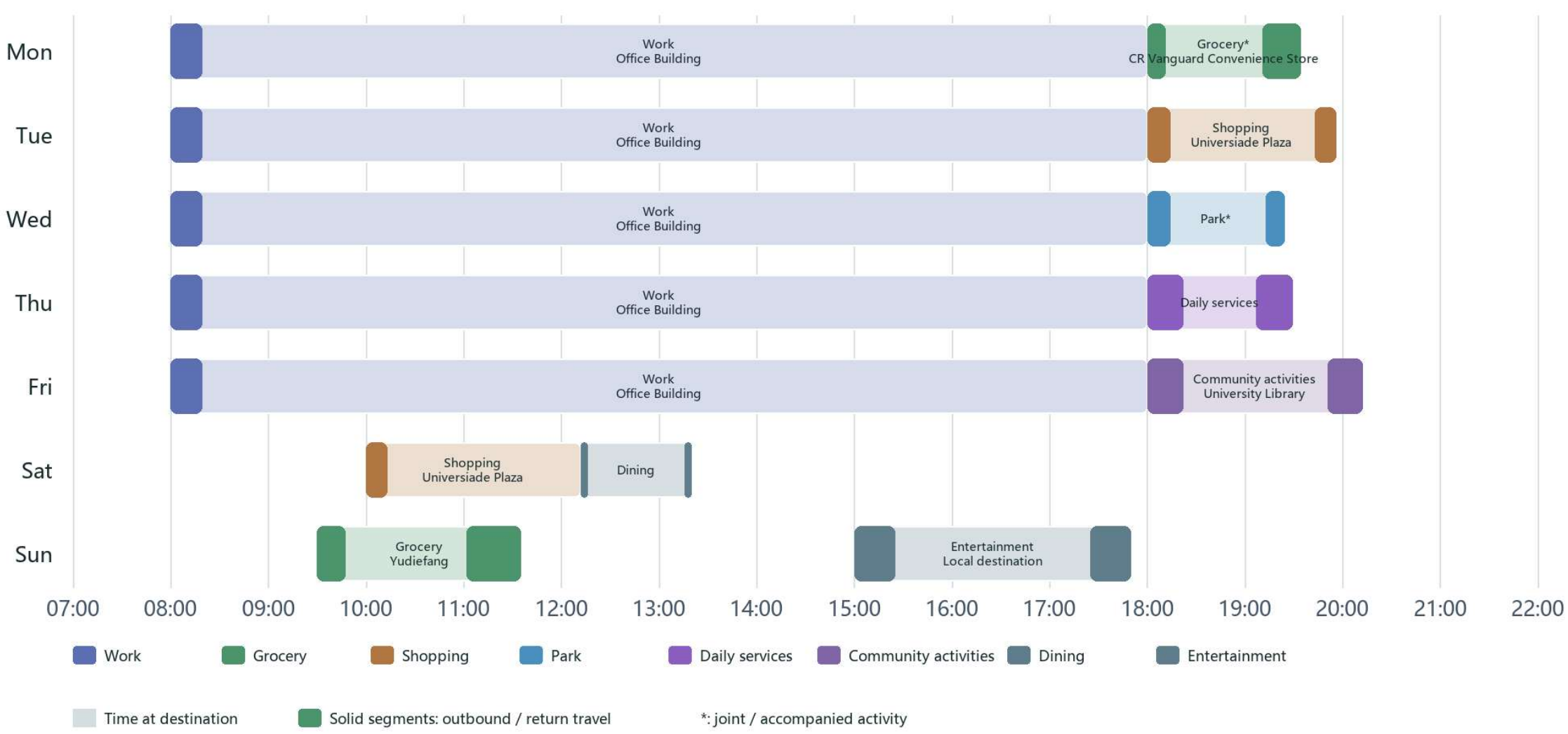
Xiao Hong's simulated weekly out-of-home activity schedule
Mon
Tue
Wed
Thu
Fri
Sat
Sun
Work
Office Building
Grocery*
CR Vanguard Convenience Store
Shopping
Universiade Plaza
Park*
Daily services
Community activities
University Library
Dining
Grocery
Yudiefang
Entertainment
Local destination
07:00
08:00
09:00
10:00
11:00
12:00
13:00
14:00
15:00
16:00
17:00
18:00
19:00
20:00
21:00
22:00
Work
Grocery
Shopping
Park
Daily services
Community activities
Dining
Entertainment
Time at destination
Solid segments: outbound / return travel
*: joint / accompanied activity

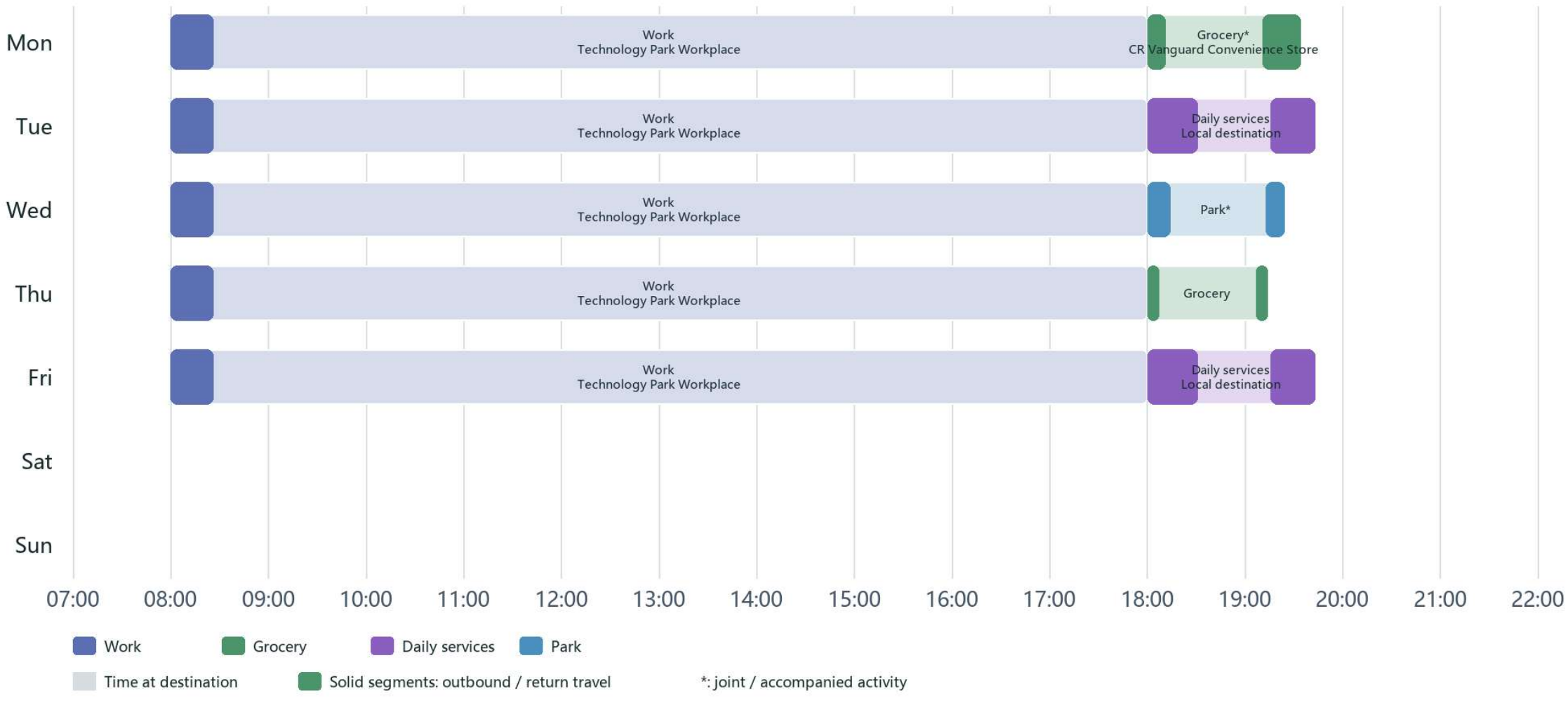
Xiao Ming's simulated weekly out-of-home activity schedule
Mon
Tue
Wed
Thu
Fri
Sat
Sun
Work
Technology Park Workplace
Grocery*
CR Vanguard Convenience Store
Daily services
Local destination
Park*
Grocery
07:00
08:00
09:00
10:00
11:00
12:00
13:00
14:00
15:00
16:00
17:00
18:00
19:00
20:00
21:00
22:00
Work
Grocery
Daily services
Park
Time at destination
Solid segments: outbound / return travel
*: joint / accompanied activity